\documentclass[prb,twocolumn,superscriptaddress,showpacs,amsmath,amssymb]{revtex4}
\usepackage{graphicx}
\usepackage{bm}
\usepackage{mathtools}

\begin{document}

\title{
Multi-band effects on Fulde-Ferrell-Larkin-Ovchinnikov states of Pauli-limited superconductors
}

\author{Masahiro Takahashi}
\email{masahiro.takahashi@gakushuin.ac.jp}
\affiliation{Department of Physics, Gakushuin University, Tokyo 171-8588, Japan}
\author{Takeshi Mizushima}
\email{mizushima@mp.okayama-u.ac.jp}
\affiliation{Department of Physics, Okayama University, Okayama 700-8530, Japan} 
\author{Kazushige Machida}
\email{machida@mp.okayama-u.ac.jp}
\affiliation{Department of Physics, Okayama University,
Okayama 700-8530, Japan}
\date{\today}

\begin{abstract}
Multi-band effects on Fulde-Ferrell-Larkin-Ovchinnikov (FFLO) states of a Pauli-limiting two-band superconductor are studied theoretically, based on self-consistent calculations of the Bogoliubov-de Gennes equation. First, we examine the phase diagrams of two-band systems with a passive band in which the intraband pairing interaction is absent and superconductivity is induced by a Cooper pair tunneling from an active band. It is demonstrated that the temperature of the Lifshitz point at which three second-order transition lines meet is independent of the Cooper pair tunneling strength. The BCS-FFLO critical field becomes lower than the Lifshitz point with increasing the interband tunneling strength, and the resultant phase diagram is qualitatively different from that in a single-band superconductor. We also study the thermodynamics of Pauli-limiting two-band superconductors with comparable intraband pairing interactions. As a consequence of a competing effect between two bands, the FFLO phase is divided into two phases: $Q_1$- and $Q_2$-FFLO phases. The $Q_1$-FFLO is favored in a high field regime and the $Q_2$-FFLO becomes stable in the lower field. In a particular case, the latter is further subdivided into a family of FFLO states with rational modulation lengths, leading to a devil's staircase structure in the field-dependence of physical quantities. The critical field, above which the FFLO is stabilized, is lower than that in a single band superconductor, while the temperature of tricritical Lifshitz point is invariant under the change of two-band parameters. 

\end{abstract}



\maketitle

\section{Introduction}
There have been intensive interests in the realization of Fulde-Ferrell-Larkin-Ovchinnikov (FFLO) states in superconductors under a high magnetic field, since the first theoretical prediction.~\cite{ff,lo} Studies on FFLO states have been expanded into various fields, ranging from condensed matter~\cite{review_sc} and cold atoms~\cite{zwierlein,liao,TM2005,TM2007,review_atom,Cai} to neutron stars.~\cite{review_nuetron} Although several candidates for FFLO superconductors have already been proposed by numerous works, direct evidence of FFLO states has not been observed in experiments. 

In a Pauli-limited superconductor, the transition from the Bardeen-Cooper-Schrieffer (BCS) state to the FFLO state is considered as the second-order transition accompanied by the Jackiw-Rebbi soliton~\cite{jackiw} which neatly accommodates Pauli paramagnetic moment.~\cite{machida1984} The soliton provides a generic key concept for commonly understanding the essential physics of FFLO phases in a single-band superconductor, incommensurate structures,~\cite{su,takayama} and fermionic excitations bound at topological defects of a superconductor through $\pi$-phase shift of the background fields.~\cite{TM2005v2}

Multi-band effects originating from multiple sheets of Fermi surfaces have recently been observed in a variety of superconductors, such as MgB$_2$, iron-based superconductors, and Sr$_2$RuO$_4$.~\cite{komendova,review214} Among them, strong Pauli effects have been reported in iron-pnictides.~\cite{cho,terashima,burger,terashima2010,terashima2013,zocco,kittaka} It is now recognized that multi-band effects are crucial to understand unconventional superconductivity. Nevertheless, in contrast to extensive studies in single-band superconductors,~\cite{review_sc} the structure and thermodynamic stability of FFLO phases in multi-band systems have not been clarified so far other than a few exception.~\cite{gurevich,ptok,Mizushima_PRL} 

The Pauli-limiting field $H_\mathrm{p}$ for a single-band system is given by $\mu_{\mathrm{B}}H^{\mathrm{single}}_{\mathrm{p}} = \Delta _0/\sqrt{2}$, which depends only on the gap, $\Delta_0$, at $T = 0$ and $H=0$. This formula is easily extended to a two-band case as 
\begin{equation}
\mu_{\mathrm{B}}H^{\mathrm{two}}_\mathrm{p} 
= \sqrt{\frac{1}{2} \left( 
\frac{\mathcal{N}_\mathrm{F}^{(1)}}{\mathcal{N}_\mathrm{F}} \left [ \Delta_{0}^{(1)} \right ]^2 
+ \frac{\mathcal{N}_\mathrm{F}^{(2)}}{\mathcal{N}_\mathrm{F}} \left [ \Delta_{0}^{(2)} \right ]^2 
\right)},
\label{eq:pauli}
\end{equation}
where $\mathcal{N}_\mathrm{F}^{(\gamma)}$ ($\Delta_0^{(\gamma)}$) is the density of states (gap) for the $\gamma$~band ($\gamma = 1$ or $2$) and we introduce $\mathcal{N}_\mathrm{F} = \mathcal{N}_\mathrm{F}^{(1)} + \mathcal{N}_\mathrm{F}^{(2)}$. When the major gap $\Delta_0^{(1)}$ dominates the minor gap $\Delta_0^{(2)}$ ($\Delta_0^{(1)} \gg \Delta_0^{(2)}$), the Pauli-limiting field in the case of two-band systems, 
$\mu_\mathrm{B}H^\mathrm{two}_\mathrm{p} \simeq \sqrt{\mathcal{N}_\mathrm{F}^{(1)} / \mathcal{N}_\mathrm{F}} \left [ \Delta_0^{(1)} / \sqrt{2} \right ]$, 
is smaller than 
$\mu_\mathrm{B}H^\mathrm{single}_\mathrm{p} = \Delta_0^{(1)} / \sqrt{2}$. 
For $\mathcal{N}_\mathrm{F}^{(2)} \gg \mathcal{N}_\mathrm{F}^{(1)}$, Eq.~\eqref{eq:pauli} reduces to
$\mu_\mathrm{B}H^\mathrm{two}_\mathrm{p} \simeq \Delta_0^{(2)} / \sqrt{2} = \mu_\mathrm{B} H^{(2)}_\mathrm{p} < \mu_\mathrm{B} H^{(1)}_\mathrm{p}$ 
with $H^{(\gamma)}_\mathrm{p}$ being the Pauli-limiting field for the $\gamma$ band. 
These two limiting cases show that the FFLO region could become enlarged towards the lower field. This simple consideration motivates us to look into the Pauli paramagnetic effects in multi-band superconductivity.

The FFLO states are Cooper pairing states with a finite center-of-mass momentum $Q$. The nonzero center-of-mass momentum generates a spatial modulation of Cooper pair amplitudes in real space, \textit{e.g.}, $\Delta (z) \propto \sin(Qz)$. The modulation period 
$2\pi/Q$
is determined by competing effect between spin paramagnetism and superconductivity. For a single-band case, the analytic form of the modulation length is derived in Ref.~\onlinecite{machida1984}. At zero temperature, the modulation length depends on the amplitude of the pair potential at zero field, the Fermi wavelength of background electrons, and an external field. In the case of a two-band superconductor, each band has its own favorable value of the modulation 
period, $2\pi/Q^{(1)}$ and $2\pi/Q^{(2)}$. 
The competing and synergetic effect arises from a coherent coupling of two bands having two modulation 
periods $2\pi/Q^{(1)}$ and $2\pi/Q^{(2)}$ 
through interband Cooper pair tunneling, which makes the phase diagram of FFLO states rich and complex. We also note that multi-band effects might be accompanied by exotic superconductivity.~\cite{garaud,babaev1,babaev2,kogan,hirano}

In the previous work,~\cite{Mizushima_PRL} it is found that in a Pauli-limiting two-band superconductor, the coherent coupling of two bands through the interband Cooper pair tunneling gives rise to a devil's staircase structure in the field dependence of physical quantities and a rich FFLO phase diagram. In this paper, we more extensively investigate the thermodynamic stability and electronic structures of Pauli-limiting two-band superconductors in a wide region of parameters. Based on self-consistent calculation of the Bogoliubov-de Gennes (BdG) equation, we first start to examine the phase diagram of a simplest two-band superconductor with a passive band in which the pairing interaction is absent and the superconductivity is induced by the Cooper pairing transfer from the active band. The resultant phase diagram is accompanied by the positive slope of the BCS-FFLO phase boundary line, $T_{\mathrm{LO}}(H)$, 
\begin{equation}
\frac{dT_\mathrm{LO}(H)}{dH} > 0.
\end{equation} 
This is contrast to the phase boundary in a single-band superconductor, ${dT_\mathrm{LO}}/{dH} < 0$ (see the solid line of Fig.~\ref{fig:phase_single} and Refs.~\onlinecite{review_sc} and \onlinecite{suzuki}). We investigate the phase diagram in the more general case that two bands with comparable intraband interactions are coherently coupled through the Cooper pair tunneling. It is demonstrated that the sequence of FFLO phases with rational modulation length is widely observed when the tunneling rate is present. For all parameters used in the current work, the tricritical Lifshitz point ``L" is found to be a fixed point, $T_\mathrm{L}/T_\mathrm{c} = 0.56\cdots$. 

This article is organized as follows. In the next section, we introduce the BdG equation for a multi-band system. The details of the derivation are supplemented in Appendix~\ref{sec:bdgformalism}. In Appendix~\ref{sec:B}, we also describe the way to numerically solve the BdG equation with a periodic boundary condition. The analytic form for the phase boundaries between superconducting states and normal states is derived in the latter half of the section. In Sec.~\ref{sec:single}, we consider a two-band superconductor with a passive band, where the interaction for intraband Cooper pairing formation is absent in the second (passive) band and the superconductivity is induced by the Cooper pair tunneling from the first (active) band. In Sec.~\ref{sec:2bands}, we show the numerical results on FFLO states of Pauli-limiting two-band superconductors, where both the bands have intraband pairing interaction. Finally, Sec.~\ref{sec:summary} is devoted to summary.

\section{Formulation}

\subsection{Multi-band Bogoliubov-de Gennes equation}

We start from the second-quantized microscopic Hamiltonian with multiple bands under a uniform external magnetic field. 
The Hamiltonian is given as
\begin{equation}
\mathcal{H} = \mathcal{H}_0 + \mathcal{H}_\mathrm{int}, 
\label{eq:Hamiltonian} 
\end{equation}
where the single-particle Hamiltonian and interaction Hamiltonian are
\begin{gather}
\mathcal{H}_0 = \sum_{\gamma, \sigma, \sigma'}\int d\bm{r}\psi_{\sigma}^{(\gamma)\dagger} (\bm{r}) \underline{\xi}_{\sigma, \sigma'}^{(\gamma)} (\bm{r}) \psi_{\sigma'}^{(\gamma)} (\bm{r}), \\
\mathcal{H}_\mathrm{int} = \sum_{\gamma, \gamma'} g_{\gamma \gamma'} \int d\bm{r}
\psi_{\uparrow}^{(\gamma)\dagger} (\bm{r}) \psi_{\downarrow}^{(\gamma)\dagger} (\bm{r}) \psi_{\downarrow}^{(\gamma')} (\bm{r}) \psi_{\uparrow}^{(\gamma')} (\bm{r}).
\label{eq:int_part}
\end{gather} 
The sums of $\sigma, \sigma'$ are taken for up- and down-spins $\uparrow, \downarrow$, and electron bands are denoted by $\gamma = 1, 2, \dots, N$.
The single-particle Hamiltonian density $\underline{\xi}_{\sigma, \sigma'}^{(\gamma)}$ is a $(\sigma, \sigma')$ component of a two-dimensional matrix $\underline{\xi}^{(\gamma)}$,
\begin{equation}
\underline{\xi}^{(\gamma)} (\bm{r}) = \varepsilon^{(\gamma)} (\bm{r}) \underline{1}_2 - \mu_\mathrm{B}^{(\gamma)} \bm{H} \cdot \underline{\bm{\sigma}},
\label{eq:HamiltonianDensity} 
\end{equation}
with $\varepsilon^{(\gamma)} (\bm{r}) = \frac{\hbar^2}{2 M_\mathrm{e}^{(\gamma)}} \left ( - i \nabla + \bm{A} \right )^2 - \mu_\gamma$. The creation and annihilation field operators, $\psi_\sigma^{(\gamma)\dagger}$ and $\psi_\sigma^{(\gamma)}$, are defined for each band and spin state. The electron mass and effective Bohr magneton in the $\gamma$ band are denoted as $M_\mathrm{e}^{(\gamma)}$ and $\mu_\mathrm{B}^{(\gamma)}$, respectively. The $n$-dimensional unit matrix is denoted as $\underline{1}_n$. We assume the effective coupling of electrons in the bands $\gamma$ and $\gamma^{\prime}$ as $g_{\gamma\gamma^{\prime}} = g_{\gamma^{\prime}\gamma} (\le 0)$. Here $\bm{A}$ is a vector potential, $\mu_\gamma$ is a chemical potential of the $\gamma$ band, and $\bm{H}$ is an external magnetic field coupled with spins through $\underline{\bm{\sigma}} = (\underline{\sigma}_x, \underline{\sigma}_y, \underline{\sigma}_z)$, where $\underline{\sigma}_\alpha~(\alpha = x, y, z)$ are $2 \times 2$ Pauli matrices.

Employing the mean field approximation and following the usual procedure,~\cite{suhl} we have the following BdG equation for the quasiparticle wavefunction $\bm{\varphi}_\nu^{(\gamma)} (\bm{r}) = [ u_\nu^{(\gamma)} (\bm{r}), v_\nu^{(\gamma)} (\bm{r}) ]^\mathrm{T}$,
\begin{gather}
\underline{\mathcal{K}}^{(\gamma)} (\bm{r}) \bm{\varphi}_{\nu}^{(\gamma)} (\bm{r}) = E_\nu^{(\gamma)} \bm{\varphi}_{\nu}^{(\gamma)} (\bm{r}), 
\label{eq:BdGfinal} 
\end{gather}
where the $2 \times 2$ BdG matrix is 
\begin{gather}
\underline{\mathcal{K}}^{(\gamma)} (\bm{r}) 
= \begin{bmatrix}
\varepsilon^{(\gamma)} (\bm{r}) - \mu_\mathrm{B}^{(\gamma)} H & \Delta^{(\gamma)} (\bm{r}) \\
\Delta^{(\gamma)\ast} (\bm{r}) & -\varepsilon^{(\gamma)} (\bm{r}) - \mu_\mathrm{B}^{(\gamma)} H \end{bmatrix}.
\label{eq:matrixK} 
\end{gather}
Equation~\eqref{eq:BdGfinal} is self-consistently coupled with the pair potential,
\begin{equation}
\Delta^{(\gamma)} (\bm{r}) = \sum_{\gamma'} \sum_{|E_\nu^{(\gamma')}| < E_\mathrm{c}} g_{\gamma \gamma'} u_\nu^{(\gamma')} (\bm{r}) v_\nu^{(\gamma') \ast} (\bm{r}) f (E_\nu^{(\gamma')}),
\label{eq:gapfinal}
\end{equation}
where $f (E) = 1/ \left ( e^{E/k_\mathrm{B}T} + 1 \right )$ is the Fermi distribution function at a temperature $T$. 
Here, we assume that $\bm{H}$ is fixed to the $z$ axis without loss of generality. The detailed formalism is explained in Appendix~\ref{sec:bdgformalism}.

In this formalism, we take account of multi-band effects through the gap equation \eqref{eq:gapfinal}. The coupling constants, $g_{\gamma\gamma}$, lead to the formation of intraband Cooper pairs, and $g_{\gamma\gamma^{\prime}}$ ($\gamma \neq \gamma^{\prime}$) induces the interband Cooper pair tunneling. Hence, in the case of nonzero $g_{\gamma\gamma^{\prime}}$ ($\gamma \neq \gamma^{\prime}$), multiple electron bands are coherently coupled with each other through the interband Cooper pair transfer. This multi-band effect is crucial for the sequence of FFLO phase transitions discussed in Sec.~\ref{sec:2bands}.

For simplicity and to stay in a minimal model, we consider a one-dimensional system along the $z$ direction. Throughout this paper, the effective masses $M_\mathrm{e}^{(\gamma)}$, and the effective Bohr magnetons $\mu_\mathrm{B}^{(\gamma)}$ of electrons in each band are set to be the same, $M_\mathrm{e} \equiv M^{(1)}_\mathrm{e} = M^{(2)}_\mathrm{e}$ and $\mu_\mathrm{B} = \mu^{(1)}_\mathrm{B} = \mu^{(2)}_\mathrm{B}$, respectively. Since we are interested in the strong Pauli paramagnetic limit, the vector potential $\bm{A}$ is supposed to be spatially uniform, neglecting the orbital repairing effect.~\cite{suzuki} Note that the orbital depairing effect can be suppressed in low-dimensional superconductors.~\cite{yonezawa,bergk}

We numerically solve the self-consistent equations \eqref{eq:BdGfinal} and \eqref{eq:gapfinal}. We consider a one-dimensional modulation of the pair potential with the period $L$,
\begin{equation}
\Delta^{(\gamma)} (z + L/2) = e^{i \chi} \Delta^{(\gamma)} (z),
\label{eq:delta_period}
\end{equation}
where $\chi = \pi ~(2\pi)$ corresponds to FFLO (BCS) states. The period $L$ determines the FFLO modulation wavenumber $Q$, 
\begin{equation}
Q \equiv 2\pi/L.
\end{equation}
This imposes the periodic boundary condition on the quasiparticle wave functions, 
\begin{equation}
\bm{\varphi}_\nu^{(\gamma)} (z + p L/2) = e^{i k R} e^{i \chi \sigma_z / 2} \bm{\varphi}_\nu^{(\gamma)} (z),
\label{eq:uv_period}
\end{equation}
where $k = \frac{2\pi q}{L N_L}$ is the Bloch vector and $R$ denotes the Bravais lattice vector $R = p L /2$ in the one-dimensional system, which satisfies $k R = \pi pq/N_L$ with $p, q, N_L \in \mathbb{Z}$. We numerically diagonalize the BdG equation \eqref{eq:BdGfinal} with the periodic boundary condition by using an finite element method implemented with the Gauss-Lobatto discrete variable representation. The details are described in Appendix~\ref{sec:B}.

Using the self-consistent solutions, we evaluate the thermodynamic potential $\Omega \equiv \langle \mathcal{H}_\mathrm{MF} \rangle - T S$, 
\begin{align}
\Omega =& E_\mathrm{cond} + \sum_\gamma \sum_{|E_\nu^{(\gamma)}| < E_\mathrm{c}} \Bigl [ E_\nu^{(\gamma)} f(E_\nu^{(\gamma)}) 
\int d\bm{r}\left|u^{(\gamma)}_{\nu}(\bm{r})\right|^2 \nonumber \\
&- k_\mathrm{B} T \ln \left ( 1 + e^{-E_\nu^{(\gamma)} / k_\mathrm{B} T} \right ) \Bigr ].
\label{eq:thermodynamicpot}
\end{align}
where 
$E_\mathrm{cond} \!=\! - \sum _{\gamma,\gamma^{\prime}} g_{\gamma\gamma^{\prime}} \int d\bm{r}\Phi^{\ast}_{\gamma}(\bm{r})\Phi _{\gamma^{\prime}}(\bm{r})$, 
with $\Phi _{\gamma}(\bm{r}) \!\equiv\! \sum _{|E_\nu^{(\gamma)}| < E_\mathrm{c}}u^{(\gamma)}_{\nu}(\bm{r})v^{(\gamma)\ast}_{\nu}(\bm{r})f(E^{(\gamma)}_{\nu})$.
The thermodynamic potential in Eq.~\eqref{eq:thermodynamicpot} is calculated for both BCS and FFLO states under a fixed $H$ and $T$. And the modulation period $L$ in equilibrium is determined so as to minimize $\Omega(Q)$ with respect to the FFLO modulation $Q$.
The other physical quantities are also calculated. The magnetization density is defined as
\begin{equation}
M (z) = M^{(1)} (z) + {M}^{(2)} (z),
\end{equation}
where we introduce the magnetization in the $\gamma$ band as
\begin{align}
{M}^{(\gamma)} (z) = \sum_{|E_\nu^{(\gamma)}| < E_\mathrm{c}} &\Bigl [ |u_\nu^{(\gamma)} (z) |^2 f (E_\nu^{(\gamma)}) \nonumber \\
&- |v_\nu^{(\gamma)} (z) |^2 f(-E_\nu^{(\gamma)}) \Bigr ].
\end{align}

\subsection{Analytic derivation of critical field $H_{c2}$}
The boundary between normal and superconducting states, $H_{c2}(T)$, can be analytically derived in multi-band systems. The FF and LO states are energetically degenerate on the boundary. Therefore, to determine the $H_{c2}(T)$ line, we assume the pair potential $\Delta^{(\gamma)}(\bm{r})$ to be the FF state, which is given by 
\begin{equation}
\Delta^{(\gamma)} (z) = \tilde{\Delta}^{(\gamma)} e^{i Q z},
\label{eq:delta_FF}
\end{equation}
where $\tilde{\Delta}^{(\gamma)} = \tilde{\Delta}^{(\gamma)} (H,T) ~(\in \mathbb{R})$ is a constant and $Q$ denotes the modulation wavenumber of the FF state.

The BdG equation \eqref{eq:BdGfinal} can be transformed to the particle-hole symmetrized form,
\begin{equation}
\underline{K}^{(\gamma)} (z) \bm{\varphi}_\nu^{(\gamma)} (z) 
= \tilde{E}_\nu^{(\gamma)} \bm{\varphi}_\nu^{(\gamma)} (z), 
\label{eq:BdGequationfinal} 
\end{equation}
where $\tilde{E}_{\nu}^{(\gamma)} = E_{\nu}^{(\gamma)} + \mu_\mathrm{B}^{(\gamma)} H$ and
\begin{gather}
\underline{K}^{(\gamma)} (z) =\varepsilon^{(\gamma)} (z)\underline{\sigma}_z + \underline{\sigma}_x e^{-iQz\underline{\sigma}_z}\tilde{\Delta}^{(\gamma)} .
\label{eq:BdGmatrixsym} 
\end{gather}
The symmetrized BdG matrix \eqref{eq:BdGmatrixsym} satisfies $\underline{\mathcal{C}} \underline{K}^{(\gamma)} (z) \underline{\mathcal{C}}^{-1} = - \underline{K}^{(\gamma)} (z)$ with $\underline{\mathcal{C}} = - i\underline{\sigma}_y K$, where $K$ is the complex conjugate operator. This implies that the positive energy solution $\tilde{E}_\nu^{(\gamma)} > 0$ with $[ u_\nu^{(\gamma)} (z), v_\nu^{(\gamma)} (z) ]^\mathrm{T}$ in Eq.~\eqref{eq:BdGequationfinal} is mapped to the negative energy solution $\tilde{E}_\nu^{(\gamma)} < 0$ with $\underline{\mathcal{C}}\bm{\varphi}_\nu^{(\gamma)} (z)$. Then the gap equation \eqref{eq:gapfinal} is reduced to 
\begin{align}
\Delta^{(\gamma)} (z) \simeq& \sum_{\gamma'} g_{\gamma \gamma'} \sum_{0 < \tilde{E}_\nu^{(\gamma')}<E_\mathrm{c}} u_\nu^{(\gamma')} (z) v_\nu^{(\gamma')\ast} (z) \nonumber \\
& \times \left [ f(\tilde{E}_\nu^{(\gamma')} - \mu_\mathrm{B} H) - f(-\tilde{E}_\nu^{(\gamma')} - \mu_\mathrm{B} H) \right ],
\label{eq:gapequationpositive}
\end{align}
where $E_\mathrm{c} \gg \mu_\mathrm{B}H$ is considered for the upper bound for the summation.

We carry out the $U(1)$ transformation of the quasiparticle wavefunction as
$\tilde{\bm{\varphi}}_\nu^{(\gamma)} (z) 
=  [ \tilde{u}_\nu^{(\gamma)} (z), \tilde{v}_\nu^{(\gamma)} (z)  ]^\mathrm{T} 
= \underline{U} (Q) \bm{\varphi}_\nu^{(\gamma)} (z)
$,
where
$\underline{U} (Q) = \exp \left [ - i (Q/2) z \underline{\sigma}_z \right ]$. 
The $U(1)$ transformation reduces the BdG matrix \eqref{eq:BdGmatrixsym} to 
$\tilde{\underline{K}}^{(\gamma)} (z, Q) 
=\underline{U} (Q) \underline{K}^{(\gamma)} (z) \underline{U}^\dagger (Q) $.
The transformed BdG matrix satisfies
$ [ \underline{\tilde{K}}^{(\gamma)} (z,Q), -i\frac{d}{dz} \underline{1}_2  ] = \underline{0}_2$, 
where $\underline{0}_n$ is a $n$-dimensional zero matrix. The operators $\underline{\tilde{K}}^{(\gamma)} (z,Q)$ and $-i\frac{d}{dz}\underline{1}_2$ have the simultaneous eigenstates. 
The eigenstates can be expressed as
$\tilde{\bm{\varphi}}_k^{(\gamma)} (z) = \exp (ikz) \bm{\varphi}_k^{(\gamma)}/\sqrt{Z}$,
where $Z = \int dz$ is constant, and $\bm{\varphi}_k^{(\gamma)} = [ u_k^{(\gamma)}, v_k^{(\gamma)} ]^\mathrm{T}$ satisfies the normalization condition, $| u_k^{(\gamma)} |^2 +  | v_k^{(\gamma)}  |^2 = 1$. Then, Eq.~\eqref{eq:BdGequationfinal} is recast into
\begin{equation}
\underline{K}_Q^{(\gamma)} (k) \bm{\varphi}_k = \tilde{E}^{(\gamma)} (k,Q) \bm{\varphi}_k, 
\end{equation}
where 
\begin{equation}
\underline{K}^{(\gamma)}_Q (k) = \begin{bmatrix} \varepsilon_Q^{(\gamma)} (k)  & \tilde{\Delta}^{(\gamma)} \\
\tilde{\Delta}^{(\gamma)} & -\varepsilon^{(\gamma)}_{-Q} (k) \end{bmatrix}, 
\end{equation}
with
$\varepsilon^{(\gamma)}_Q (k) = \frac{\hbar^2}{2M_\mathrm{e}}  (k + \frac{Q}{2} )^2 - \mu_\gamma$.
Finally, the eigenenergy $\tilde{E}^{(\gamma)} (k,Q)$ is calculated as
\begin{equation}
\tilde{E}^{(\gamma)} (k,Q) = \frac{\hbar^2}{2M_\mathrm{e}} k Q
\pm \mathcal{E}^{(\gamma)} (k,Q), 
\label{eq:eigenenergy}
\end{equation}
with 
\begin{equation}
\mathcal{E}^{(\gamma)} (k,Q) = \sqrt{\varepsilon^{(\gamma)}_Q (k) \varepsilon^{(\gamma)}_{-Q} (k) + \left (\frac{\hbar^2kQ}{2M_\mathrm{e}} \right )^2 + \left [ \tilde{\Delta}^{(\gamma)} \right ]^2 }. 
\end{equation}
The corresponding eigenstates are given as
\begin{align}
u_k^{(\gamma)} =& \sqrt{\frac{1}{2} \left [ 1 + \frac{\varepsilon^{(\gamma)}_{-Q} (k) + \frac{\hbar^2}{2M_\mathrm{e}} kQ}{\tilde{E}^{(\gamma)} (k,Q) - \frac{\hbar^2}{2M_\mathrm{e}} kQ} \right ]}, 
\label{eq:u_k} \\
v_k^{(\gamma)} =& \sqrt{\frac{1}{2} \left [ 1 - \frac{\varepsilon^{(\gamma)}_Q (k) - \frac{\hbar^2}{2M_\mathrm{e}} kQ}{\tilde{E}^{(\gamma)} (k,Q) - \frac{\hbar^2}{2M_\mathrm{e}} kQ} \right ]}. 
\label{eq:v_k}
\end{align}

Using these eigenenergies and eigenfunctions, the pair potential \eqref{eq:gapequationpositive} becomes 
\begin{gather}
\Delta^{(\gamma)} (z) = - \sum_{\gamma'} g_{\gamma \gamma'} {\sum_{\tilde{E}^{(\gamma')}}}^{+} 
~\frac{\Delta^{(\gamma')} (z)}{2 Z} \mathcal{F}^{(\gamma')}(k,Q), \\
\mathcal{F}^{(\gamma)} (k,Q) = 
\frac{1 - f_{+}^{(\gamma)} - f_{-}^{(\gamma)}}
{\mathcal{E}^{(\gamma)} (k,Q)},
\end{gather}
where $f_{\pm}^{(\gamma)} = f (\tilde{E}^{(\gamma)} (k,Q) \pm \mu_\mathrm{B} H)$ and $\sum_{\tilde{E}^{(\gamma')}}^{+}$ means summation in $0 < \tilde{E}^{(\gamma')} (k,Q) < E_\mathrm{c}$. During the calculation, positive sign is taken for $\tilde{E}^{(\gamma)} (k,Q)$ in Eq.~\eqref{eq:eigenenergy}.
Then the reduced gap equation for
$
\bm{\Delta}
= [
\Delta^{(1)}, \Delta^{(2)}, \dots , \Delta^{(N)}
]^\mathrm{T}$ is expressed in the $N$-dimensional matrix form as
\begin{equation}
\left [ \underline{1}_N - \underline{\mathcal{I}}_N (Q) \right ] \bm{\Delta}
= \bm{0}, 
\end{equation}
with 
$\left[\underline{\mathcal{I}}_N (Q)\right]_{\gamma\gamma^{\prime}} = g_{\gamma\gamma^{\prime}} 
\mathcal{I}^{(\gamma^{\prime})} (Q) $ and 
$\mathcal{I}^{(\gamma)} (Q) = -\frac{1}{2} \int_{-k_\mathrm{c}}^{k_\mathrm{c}} \frac{dk}{2\pi} \mathcal{F}^{(\gamma)} (k,Q)$.
We here replace $Z^{-1} \sum_{0<E<E_\mathrm{c}}$ by $\int_{-k_\mathrm{c}}^{k_\mathrm{c}} \frac{dk}{2\pi}$ in the thermodynamic limit. To have a nontrivial set of $\bm{\Delta}$, we have the condition,
\begin{equation}
\mathrm{det} \left [ \underline{1}_N - \underline{\mathcal{I}}_N (Q) \right ] = 0.
\label{eq:analyticboudarydet}
\end{equation}
The analytical phase boundary $H_{c2}(T)$ for $N$-band system is obtained from Eq.~\eqref{eq:analyticboudarydet} for various $Q$'s.  

In the present paper, we consider the case of $N = 2$. Then, Eq.~\eqref{eq:analyticboudarydet} can be written as
\begin{align}
\MoveEqLeft[9] \left [ 1 - g_{11} \mathcal{I}^{(1)} (Q) \right ] \left [ 1 - g_{22} \mathcal{I}^{(2)} (Q) \right ] \nonumber \\
&- g_{12} g_{21} \mathcal{I}^{(1)} (Q) \mathcal{I}^{(2)} (Q) = 0.
\label{eq:analyticboundary}
\end{align}
To derive the $H_\mathrm{c2}(T)$ line, we vary $Q ~(>0)$ and $H ~(> 0)$ so as to find the set of parameters $(Q,H)$ which satisfy Eq.~\eqref{eq:analyticboundary} under a fixed temperature $T$. The highest value of $H$ gives the phase boundary $H_\mathrm{c2}(T)$ at a fixed temperature $T$. In a similar manner, we derive the BCS-normal boundary from Eq.~\eqref{eq:analyticboundary} with $Q = 0$. The point at which two boundaries meet corresponds to the tricritical point, the Lifshitz point ``L".

\section{Two-band systems with a passive band: the case of $g_{22} = 0$}
\label{sec:single}
Let us start from a simplest, yet nontrivial two-band case with a passive band. This corresponds to the case of $g_{22}=0$ and the intraband interaction in the second band is absent. In such a case, the superconductivity in the second (passive) band is induced by Cooper pair tunneling from the first (active) band through the interband interaction $g_{12} = g_{21}$. We study the phase diagram for various $g_{12}$'s. This gives a first step to understand the thermodynamic properties of FFLO phases in multi-band superconductors.

\begin{figure}[h!]
\includegraphics[width=\linewidth]{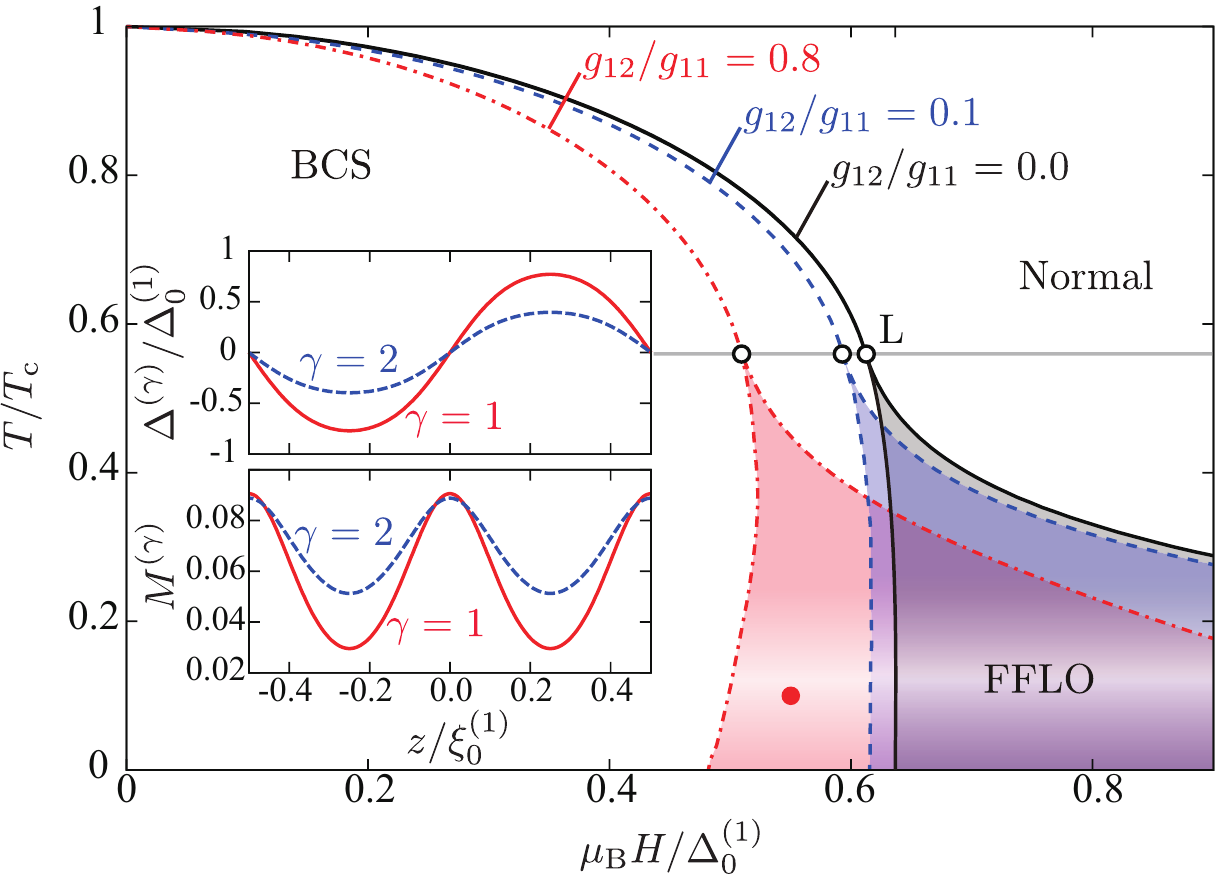}
\caption{(color online) Phase diagram for the case of two-band systems with $g_{22} = 0$. (i) The solid (black) line corresponds to the phase boundaries in a single-band superconductor ($g_{12} = g_{22} = 0$), (ii) the dashed (blue) line is the two-band case with a passive band $\bm{\Gamma} = (0.1,0.0,0.5)$ and (iii) the dash-dotted (red) line is the case of $\bm{\Gamma} = (0.8,0.0,0.5)$, where 
$\Delta_0^{(2)}/\Delta_0^{(1)} \simeq 0.10$ ($0.49$)
in the former (latter) case. The temperature of the Lifshitz point, ``L", is a fixed point $T_\mathrm{L}/T_\mathrm{c} = 0.56\cdots$, which is independent of the interband coupling $g_{12} = g_{21}$. The inset shows the spatial profile of the pair potential in the top row and the magnetization in the bottom row at $(T,H)$ marked as a filled (red) circle in the case (iii). The solid (red) lines and the dashed (blue) lines correspond to the quantities of the first and second band, respectively.}
\label{fig:phase_single}
\end{figure}

Figure~\ref{fig:phase_single} shows the phase diagram in an $H$-$T$ plane for three sets of parameters: (i) The single-band case ($g_{12} = g_{22} = 0$) and two-band cases with a passive band, (ii) $\bm{\Gamma} = (g_{12}/g_{11},g_{22}/g_{11},\mu_2/\mu_1) = (0.1,0.0,0.5)$ and (iii) $\bm{\Gamma} = (0.8,0.0,0.5)$. Using the self-consistent solutions under the parameter sets of (ii) and (iii), we estimate $\Delta_0^{(2)}/\Delta_0^{(1)}$ as $\Delta_0^{(2)}/\Delta_0^{(1)} \simeq 0.10$ and $\Delta_0^{(2)}/\Delta_0^{(1)} \simeq 0.49$, respectively. The gap amplitude of the $\gamma$ band at zero temperature and zero magnetic field is denoted by $\Delta_0^{(\gamma)}$.

The inset of Fig.~\ref{fig:phase_single} shows the spatial profile of the pair potential $\Delta^{(\gamma)}(z)$ and the magnetization density $M^{(\gamma)}(z)$ at the filled (red) circle in the main frame of Fig.~\ref{fig:phase_single} with the parameter set of the case (iii). The pair potential shows that the energy gap of the second band has a finite amplitude, implying that the Cooper pairs are transferred from the active band even if the intraband interaction in the second band is absent. The transferred Cooper pairs also affect back to the electrons in the first band through the Cooper pair tunneling from the passive band. We found the enhancement of the gap amplitude in the first band by increasing the interband interaction.

As shown in Fig.~\ref{fig:phase_single}, the phase boundary between BCS and FFLO phases shifts to lower fields as the interband interaction increases. It is important to mention that the Lifshitz point, ``L", at which three second-order phase boundaries meet shifts to the lower field as $|g_{12}|$ increases, while the temperature of the Lifshitz point, $T_\mathrm{L}$, is unchanged by the parameter $g_{12}$. 
We numerically confirm that $T_\mathrm{L}$ is invariant under any parameter changes including intraband interaction $g_{22}$. It is found that the phase boundary, $H_\mathrm{LO}(0)$, between BCS and FFLO states, at the zero temperature sharply varies as a function of the interband coupling $g_{12}$, while the magnetic field of the Lifshitz point, $H_\mathrm{L}(T_\mathrm{L})$, is relatively insensitive to $g_{12}$. This leads to the positive slope of the phase boundary, $dT_\mathrm{LO}(H) / dH > 0$, in the large interband interaction regime, where $T_\mathrm{LO}(H)$ is the BCS-FFLO transition line. This feature indicates that the interband interaction stabilizes the FFLO states even in the lower field regime. Note that all the phase boundaries in Fig.~\ref{fig:phase_single} are of the second order.

\section{Two-band systems with $g_{22}\neq 0$}
\label{sec:2bands}

In this section we set the intraband interaction in the second band to be finite, $g_{22} \neq 0$. In this case, the electrons in both the two bands essentially contribute to the superconducting state and the bands are coherently coupled to each other through the Cooper pair tunneling with $g_{12} = g_{21} \neq 0$. Before going further, let us review the analytic result of the magnetic field $H$ dependence of the FFLO wavenumber $Q$ in a single-band case.~\cite{machida1984} By introducing a parameter $k \in [0,1)$, $H$ and $Q$ are expressed with the analytic form,
\begin{gather}
h (k) \equiv \frac{\mu_\mathrm{B} H}{\Delta_0} = \frac{2}{\pi} \frac{E (k)}{k} , 
\label{eq:singleH} \\
q (k) \equiv \frac{\hbar v_\mathrm{F} Q}{\Delta_0} = \frac{\pi}{k K(k)},
\label{eq:singleQ}
\end{gather}
where $\Delta_0$ is the pairing amplitude at $(H,T) = (0,0)$, $v_\mathrm{F}$ is the Fermi velocity, and $K (k)$ and $E (k)$ are the complete elliptic functions of the first and second kinds, respectively. 

Although the equations \eqref{eq:singleH} and \eqref{eq:singleQ} are derived from a single-band Hamiltonian, it is applicable to a two-band system when the interband coupling $|g_{12}|$ is small. We here define two coherence lengths, $\xi_0^{(\gamma)}$, defined in each electron band,
\begin{equation}
\xi_0^{(\gamma)} k_\mathrm{F}^{(\gamma)} = \frac{2 \mu_\gamma}{\Delta_0^{(\gamma)}}, ~~(\gamma = 1, 2),
\label{eq:coherences}
\end{equation}
where $k_\mathrm{F}^{(\gamma)} = \sqrt{2M_\mathrm{e} \mu_\gamma/\hbar^2}$ is the Fermi wavenumber of the $\gamma$ band. Let us here introduce the averaged variables of various quantities. The averaged gap at $(H,T) = (0,0)$ is $\bar{\Delta}_0 = ( \Delta_0^{(1)} + \Delta_0^{(2)} ) / 2$, 
the averaged chemical potential is $\bar{\mu} = ( \mu_1 + \mu_2 ) / 2$,
and the averaged coherence length is defined as
\begin{equation}
\bar{\xi} = \frac{\sqrt{2\bar{\mu}}}{\bar{\Delta}_0} \sqrt{\frac{\hbar^2}{M_\mathrm{e}}},
\end{equation}
where the last equation is analogous to Eq.~\eqref{eq:coherences}.
Using these expressions, $Q^{(\gamma)}$ and $H^{(\gamma)}$ are characterized as,
\begin{gather}
\frac{\mu_\mathrm{B} H^{(\gamma)}}{\bar{\Delta}_0} = \frac{\Delta_0^{(\gamma)}}{\bar{\Delta}_0} h (k), 
\label{eq:Hnormalize} \\
Q^{(\gamma)} \bar{\xi} = \frac{\Delta_0^{(\gamma)}}{\bar{\Delta}_0} \sqrt{\frac{\bar{\mu}}{\mu_\gamma}} q (k).
\label{eq:Qnormalize}
\end{gather}
In this analysis, we can divide the parameter space into the two cases such that the ratio of the FFLO wavenumber in the high field limit denoted as $\mathcal{R}$ is bigger or smaller than $1$. 
From Eqs.~\eqref{eq:Hnormalize} and \eqref{eq:Qnormalize}, $\mathcal{R}$ is defined as
\begin{equation}
\mathcal{R} \equiv \lim_{\substack{H \rightarrow H_{c2} (0)\\(k \rightarrow 0)}} \frac{Q^{(1)}}{Q^{(2)}} = \lim_{\substack{H \rightarrow H_{c2} (0) \\(k \rightarrow 0)}} \frac{dQ^{(1)}/dH^{(1)}}{dQ^{(2)}/dH^{(2)}} = \sqrt{\frac{\mu_2}{\mu_1}},
\end{equation}
where $H_{c2} (0)$ is the critical field at $T = 0$. In the present model, $H_{c2} (0)$ diverges. 

\begin{figure}[t]
\includegraphics[width=0.8\linewidth]{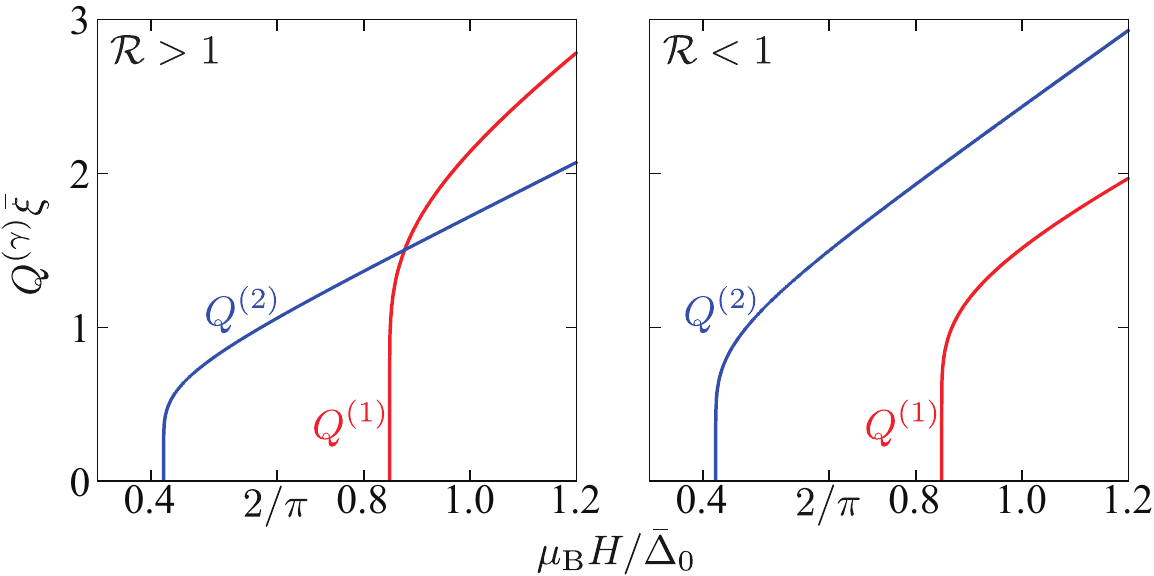}
\caption{(color online) Field-dependence of the FFLO modulation number $Q(H)$, based on the single-band analysis. The left (right) figure corresponds to the case $\mathcal{R} > 1 ~(< 1)$.}
\label{fig:analyticQs}
\end{figure}

Without loss of generality, in this paper, we consider $\Delta_0^{(1)} > \Delta_0^{(2)}$ so that $H_\mathrm{cr}^{(1)} > H_\mathrm{cr}^{(2)}$, where $H_\mathrm{cr}^{(\gamma)} = \lim_{k \rightarrow 1} H^{(\gamma)}$ in~\eqref{eq:Hnormalize}. In such a situation, if $\mathcal{R} > 1$, the curves, $Q^{(1)}(H)$ and $Q^{(2)}(H)$, are crossed at certain $H$, as shown in the left panel of Fig.~\ref{fig:analyticQs}. For the case of $\mathcal{R} < 1$, $Q^{(1)}(H)$ and $Q^{(2)}(H)$ are never crossed in all $H$'s as shown in the right panel of Fig.~\ref{fig:analyticQs}. The detailed structures on the two cases, $\mathcal{R} \gtrless 1$, are described in the 
following two subsections.

\subsection{$\mathcal{R} > 1$ case}
\label{subsec:R>1}

\subsubsection{Phase diagram}

Let us consider the $\mathcal{R} > 1$ case where the FFLO modulation wavenumbers $Q^{(\gamma)}$'s 
in the single-band analysis have a crossing point at a certain magnetic field (left panel of Fig.~\ref{fig:analyticQs}). 
Using the self-consistent solution of the BdG equation~\eqref{eq:BdGfinal} and gap equation \eqref{eq:gapfinal} with $\bm{\Gamma} = (0.1, 0.6, 2.0)$, we estimate the ratio of the order parameters at $T=0$ and $H=0$ as $\Delta_0^{(2)}/\Delta_0^{(1)} \simeq 0.31$. Then the ratio of the FFLO wavenumber is found to be $\mathcal{R} = \sqrt{2}$. 
The ground state is determined by minimizing the thermodynamic potential in Eq.~\eqref{eq:thermodynamicpot} with respect to the FFLO modulation period 
$L = 2\pi/Q$.

\begin{figure}[!t]
\includegraphics[width=\linewidth]{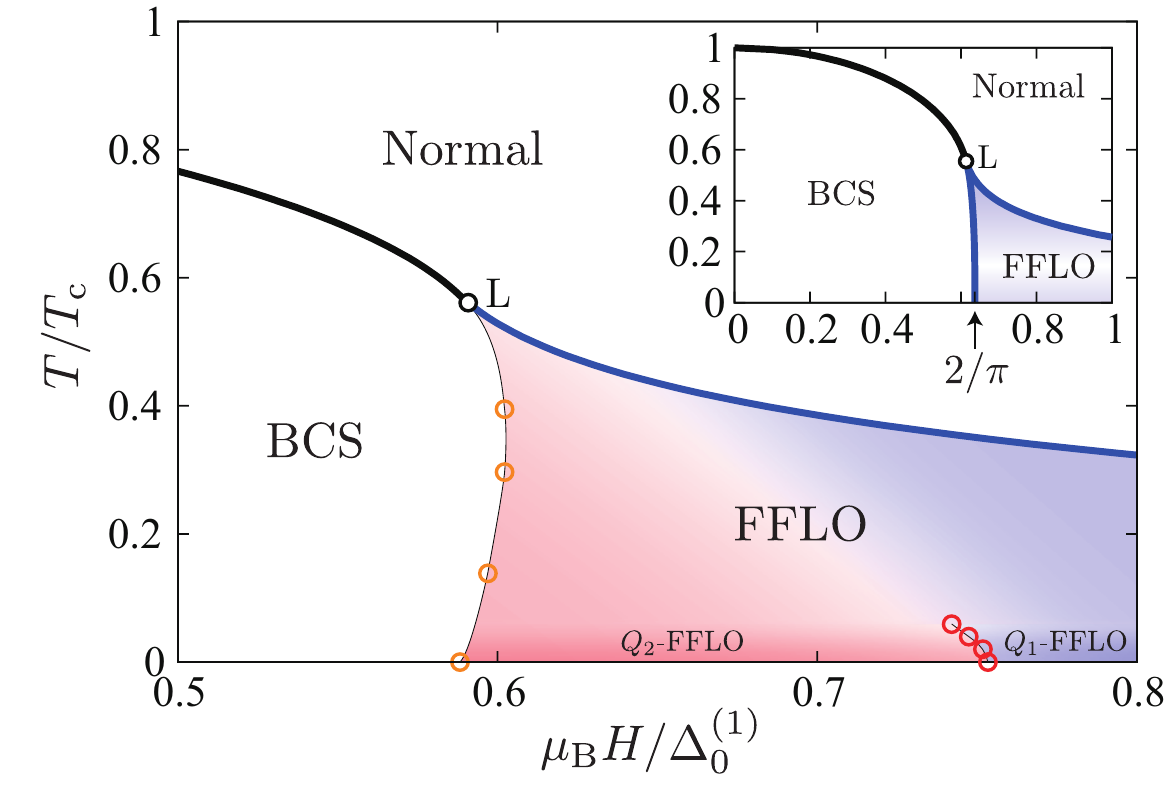}
\caption{(color online) Phase diagrams for $\bm{\Gamma} = (0.1, 0.6, 2.0)$, where $\mathcal{R}=\sqrt{2}$ and $\Delta_0^{(2)}/\Delta_0^{(1)} \simeq 0.31$. The FFLO phase is separated to two regions in the low temperature regime, $Q_1$- and $Q_2$-FFLO phases. In the high temperature regime, it is ambiguous wether the state is $Q_1$- or $Q_2$-FFLO. The inset shows the phase diagram for a single-band superconductor.}
\label{fig:phase_T-case1}
\end{figure}

The resulting phase diagram is displayed in Fig.~\ref{fig:phase_T-case1}. In the low temperature regime, the FFLO phase is divided into two regions in terms of $Q$. In Fig.~\ref{fig:phase_T-case1}, the phase in the high field regime is referred to as the $Q_1$-FFLO phase and in the lower field regime as the $Q_2$-FFLO phase. 
Since the second band has the smaller gap, 
$|\Delta^{(2)}|< |\Delta^{(1)}|$, 
the second band becomes passive in the high field regime. The $Q_1$-FFLO phase originates from the first band and its field-dependence is understandable with the single-band picture. The detailed structure will be discussed below. In the lower field regime, however, 
the gain of the condensation energy in the second band becomes more important, compared with that in the higher field. Since both bands form FFLO states with comparable amplitudes of the pair potentials, each band has the favored FFLO modulation wavenumbers as described in Eqs.~\eqref{eq:Hnormalize} and \eqref{eq:Qnormalize}. The coherent interband coupling with $g_{12}=g_{21}\neq 0$ induces frustration of two FFLO modulation wavenumbers, $Q^{(1)}$ and $Q^{(2)}$. The competing effect gives rise to emergence of new FFLO phase, that is, the $Q_2$-FFLO phase. The frustration is never seen in the case of single-band superconductors and the $Q_2$-FFLO is peculiar to multi-band systems. 
In the high temperature region, the phase boundary between the $Q_1$- and $Q_2$-FFLO phases is indistinguishable. In Fig.~\ref{fig:phase_T-case1}, the thick (thin) line indicates the second-order (first-order) phase transition. The BCS, $Q_1$-FFLO, and $Q_2$-FFLO phases are separated via the first-order phase transition, which is contrast to the second-order BCS-FFLO transition in single-band superconductors.

\begin{figure}[t]
\includegraphics[width=\linewidth]{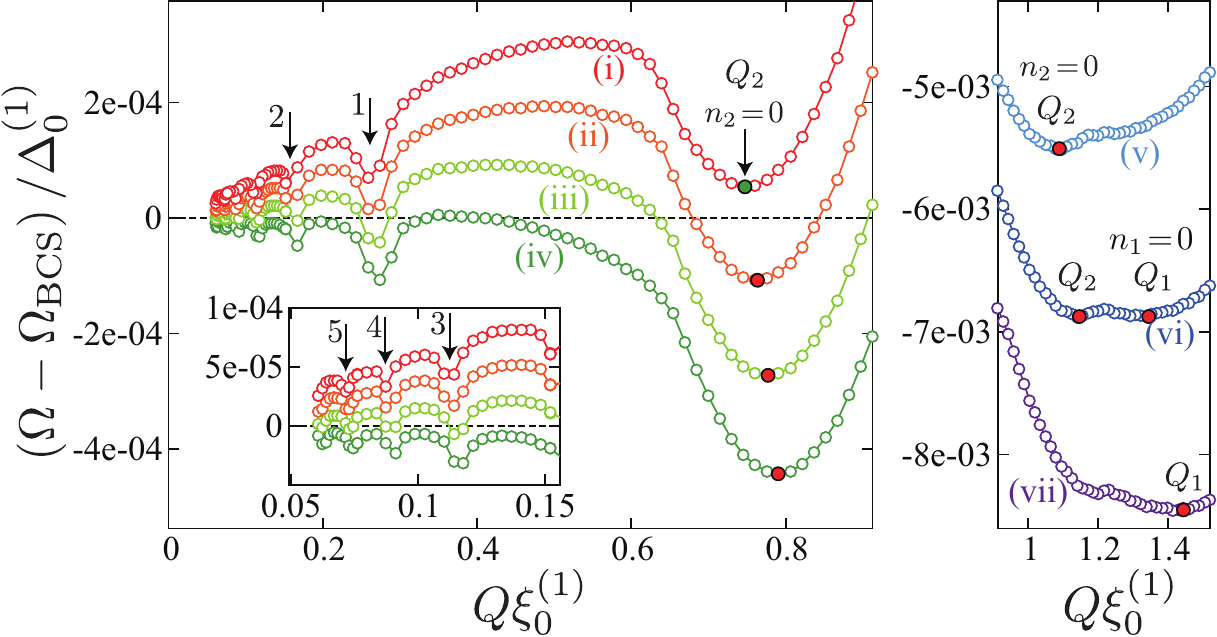}
\caption{(color online) Thermodynamic potential $\Omega$ as a function of the FFLO modulation wavenumber $Q$ at $T = 0$ for various magnetic fields, $\mu_\mathrm{B} H / \Delta_0^{(1)} = 0.586$ (i), $0.591$ (ii), $0.597$ (iii), $0.602$ (iv), $0.726$ (v), $0.753$ (vi), and $0.780$ (vii). The parameters are same as which in Fig.~\ref{fig:phase_T-case1}. The inset shows the enlarged plot of $\Omega(Q)$ in the smaller $Q$ region. The (red) filled circles show the global minima of $\Omega(Q)$.}
\label{fig:Omega-Q_T-case1}
\end{figure}

The thermodynamic potential $\Omega (Q)$ for various $H$'s is shown in Fig.~\ref{fig:Omega-Q_T-case1}. As $H$ increases, the global minimum of $\Omega (Q)$ shifts from $Q_2$-FFLO phases to the $Q_1$-FFLO phase. The $Q_1$- and $Q_2$-FFLO phases are stabilized with either the condition $Q \simeq Q^{(1)}$ or $Q \simeq Q^{(2)}$, respectively. Since $Q^{(1)} \ne Q^{(2)}$, the transition becomes the first order.

It is found that there are many local minima in $\Omega (Q)$ in Fig.~\ref{fig:Omega-Q_T-case1}. To understand the local minima, let us consider the simplified pair potential, $\Delta^{(\gamma)} (z) \propto \sin \tilde{Q}^{(\gamma)} z$. Since two bands are coherently coupled through the interband tunneling, the pair potentials have to satisfy the periodic boundary condition given by Eq.~\eqref{eq:delta_period} with $\chi = \pi$. Then $\tilde{Q}^{(1)}$ and $\tilde{Q}^{(2)}$ satisfy the condition
\begin{equation}
Q = \tilde{Q}^{(1)} / (2 n_1 + 1) = \tilde{Q}^{(2)} / (2 n_2 + 1), 
\label{eq:Q_strict}
\end{equation}
with $n_\gamma \in \mathbb{Z}$.
In Fig.~\ref{fig:Omega-Q_T-case1}, the local minima in the low $Q$'s in the low fields correspond to the cases of $n_2 = 1, 2, \dots$ and $n_1 = 0$. In the current case ($\mathcal{R} > 1$), all the states with nonzero $n_2$ are energetically unstable, relative to the $n_2=0$ $Q_2$-FFLO state.

\subsubsection{Pair potentials and local magnetization}

\begin{figure*}[!h!bt]
\begin{minipage}{0.75\linewidth}
\includegraphics[width=\linewidth]{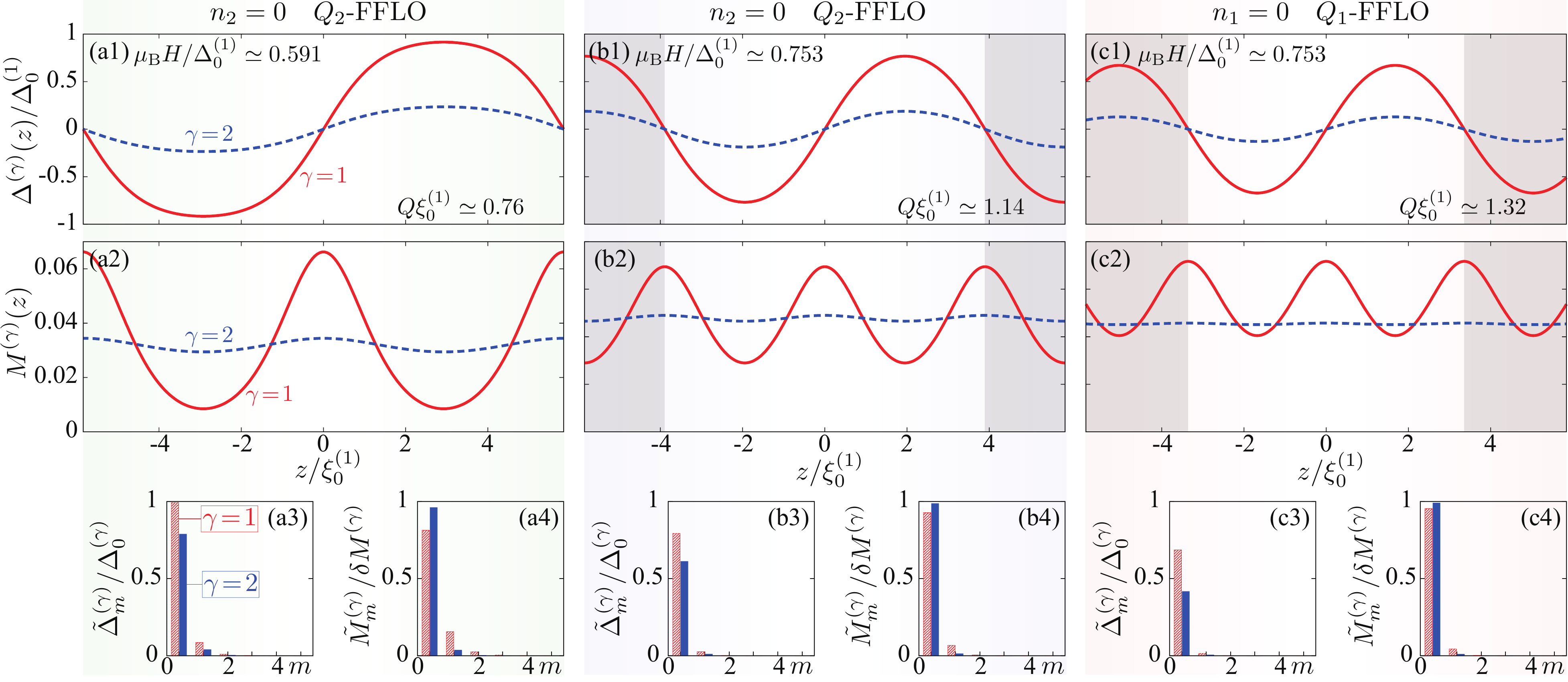}
\end{minipage}
\begin{minipage}{0.24\linewidth}
\caption{(color online) Spatial profiles of the gap for $\mu_\mathrm{B} H / \Delta_0^{(1)} = 0.591$ (a1), 
$0.753$ (b1), and $0.753$ (c1) at $T=0$. The corresponding magnetization densities are displayed in (a2), (b2), and (c2). 
(c1) and (c2) correspond to the $Q_1$-FFLO state and the others are identified as the $Q_2$-FFLO states. The parameters are same as those in Fig.~\ref{fig:phase_T-case1}. 
The bottom panels show their Fourier components where $\delta M^{(\gamma)} = \mathrm{max}[M^{(\gamma)}]-M_\mathrm{ave}^{(\gamma)}$.
}
\label{fig:delta_mag_T-case1}
\end{minipage}
\end{figure*}

Figures~\ref{fig:delta_mag_T-case1}(a1)-(c1) show the spatial profiles of the pair potentials 
and Figs.~\ref{fig:delta_mag_T-case1}(a2)-(c2) are the magnetizations
of the ground states under various $H$'s.
Here, Figs.~\ref{fig:delta_mag_T-case1}(c1) and \ref{fig:delta_mag_T-case1}(c2) correspond to the local 
minimum with the largest $Q$ at the critical field shown 
in the curve (vi) in Fig.~\ref{fig:Omega-Q_T-case1}.
The pair potential and magnetization density of the local minimum with the second largest $Q$
are displayed in (b1) and (b2) of Fig.~\ref{fig:delta_mag_T-case1}.
The data in Figs.~\ref{fig:delta_mag_T-case1}(a1) and \ref{fig:delta_mag_T-case1}(a2) 
illustrate the global-minimum state with the largest $Q$ in the curve (ii). 

To understand the spatial profiles of $\Delta^{(\gamma)}$, we here carry out the Fourier expansion of the pair potential, 
$\Delta^{(\gamma)} (z) = \sum_{m_\gamma \in \mathbb{N}} \tilde{\Delta}_{m_\gamma}^{(\gamma)} \sin \left [ q_\gamma (2 m_\gamma + 1) z \right ]$ where $\mathbb{N} = \{0, 1, 2, \dots\}$.
The periodic boundary condition~\eqref{eq:delta_period} imposes the following condition on $q_\gamma$: $Q (2 m - 1) = q_1 (2 m_1 - 1) = q_2 (2 m_2 - 1)$, where $m, m_\gamma \in \mathbb{N}$ and $Q$ is an FFLO wavenumber. Then the Fourier expansion of the pair potential is given for the first and second bands as
\begin{equation}
\Delta^{(\gamma)} (z) = 
\sum_{m \in \mathbb{N}} \tilde{\Delta}_m^{(\gamma)} 
\sin \left [ \left (2 m + 1 \right ) Q z \right ].
\end{equation}
Due to the symmetry of the pair potential in FFLO states, the Fourier components only take odd numbers, $2m +1$. The Fourier components of the pair potential is given by 
\begin{equation}
\tilde{\Delta}_{m}^{(\gamma)} = \frac{4}{L} \int_0^{L/2} 
\Delta^{(\gamma)} (z) \sin \left [ \left ( 2 m + 1 \right ) Q z \right ] dz.
\label{eq:Dm}
\end{equation}
Similarly, the magnetization density is expanded in terms of the Fourier series as
\begin{equation}
M^{(\gamma)} (z) = M_\mathrm{ave}^{(\gamma)} + \sum_{m \in \mathbb{N}} \tilde{{M}}_{m}^{(\gamma)} 
\cos \left [ 2 \left ( m + 1 \right ) Q z \right ], 
\end{equation}
with an average value of the magnetization, 
\begin{equation}
M^{(\gamma)}_\mathrm{ave} = \frac{2}{L} \int_0^{L/2} M^{(\gamma)} (z) dz.
\end{equation}
The Fourier components are given by
\begin{equation}
\tilde{M}_{m}^{(\gamma)} = \frac{4}{L} \int_0^{L/2} M^{(\gamma)} (z) 
\cos \left [ 2 \left ( m + 1 \right ) Q z \right ] dz.
\label{eq:Mm}
\end{equation}

The Fourier components defined in Eqs.~\eqref{eq:Dm} and \eqref{eq:Mm} are displayed in the bottom of Fig.~\ref{fig:delta_mag_T-case1}. From the Fourier components in Figs.~\ref{fig:delta_mag_T-case1}(a3)-(c3)
and Figs.~\ref{fig:delta_mag_T-case1}(a4)-(c4), these states are found to be purely sinusoidal states with a single modulation wavenumber $Q$. The other local minima in Fig.~\ref{fig:Omega-Q_T-case1} are classified as the $Q_2$-FFLO states with $n_2 > 0$. These states can not be the ground state in the case of $\mathcal{R}>1$, while they have the lowest energy in the case of $\mathcal{R}<1$. Hence, the detailed structures of the $Q_2$-FFLO states with $n_2 > 0$ are described in Sec.~\ref{subsec:R<1}.

The Fourier components in Fig.~\ref{fig:delta_mag_T-case1} indicate that the pair potentials and the magnetizations are well described with a sinusoidal form with a single FFLO modulation wavenumber. To understand this, let us first clarify the $Q(H)$ curve in the case of a single-band superconductor. In a single-band superconductor, the BCS-FFLO phase transition occurs at the critical field $\mu_\mathrm{B} H = (2 / \pi) \Delta_0$. The critical field is associated with the one-soliton formation energy and an isolated kink state of $\Delta (z)$ is characterized by 
$\Delta (z) = \sum_{m \in \mathbb{N}} \tilde{\Delta}_m \sin [ Q (2m +1) z ]$ 
with $\tilde{\Delta}_m = \Delta_0 / (2m + 1)$.~\cite{machida1984} The higher Fourier components with $m \ge 1$ immediately disappear as $H$ increases and the spatial modulation of $\Delta$ results in a sinusoidal form $\Delta (z) \propto \sin (Qz)$. The field dependence of $Q$ in the single-band case is described with Eqs.~\eqref{eq:singleH} and \eqref{eq:singleQ}. The slope of $Q$ with respect to $H$ in the limit of $H \rightarrow \infty$ is obtained as
\begin{equation}
\lim_{H \rightarrow \infty} \frac{\hbar v_\mathrm{F} Q  / \Delta_0}{\mu_\mathrm{B} H/\Delta_0} 
= \lim_{k\rightarrow 0} \frac{\pi/[kK(k)]}{2 E(k)/[\pi k]} = 2.
\end{equation}
This indicates $Q \sim 2 \mu_\mathrm{B} H / \hbar v_\mathrm{F}$ in the limit of the high magnetic field.

\begin{figure}[!b]
\includegraphics[width=\linewidth]{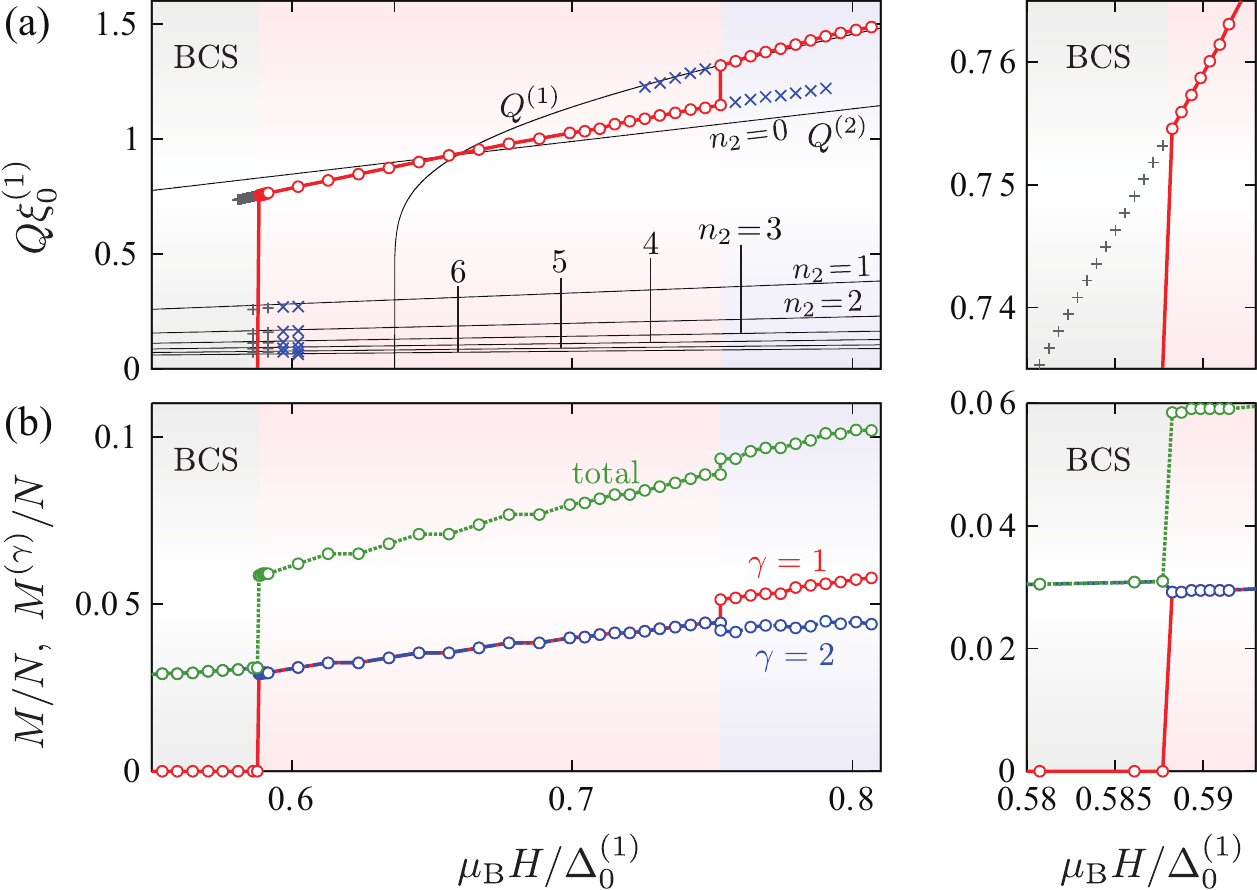}
\caption{(color online) (a) FFLO wavenumber $Q$ and (b) spatially averaged magnetization, $M$ and $M^{(\gamma)}$, as a function of $H$ at $T = 0$. The other parameters are same as those in Fig.~\ref{fig:phase_T-case1}. The (black) thin lines indicate the $Q^{(\gamma)}(H)$ curves obtained from the single-band analysis. The lines at lower $Q$ correspond to the family of the $Q_2$-FFLO, $Q^{(2)}/(2n_2+1)$, with $n_2 \in \mathbb{N}$. The symbol ``$\times$'' (``$+$'') indicates the local minimum state whose thermodynamic potential is lower (higher) than that in the BCS state. The right panels show the detailed structures in the lower $H$ region. }
\label{fig:Q_M-h_T-case1}
\end{figure}

In the case of two-band superconductors with $\Delta^{(1)}_0 > \Delta^{(2)}_0$, electrons in the second band feel the relatively high magnetic field even in the vicinity of the phase boundary between BCS to $Q_2$-FFLO phases. Therefore, the higher Fourier components never appear. Since pair potentials in two bands are coherently coupled through the interband Cooper pair tunneling, the pair potential in the first band is also pulled into a sinusoidal form by the effect of the second band.

Figure~\ref{fig:Q_M-h_T-case1} shows the field dependence of the FFLO wavenumber $Q$ and the spatially averaged magnetization $M_\mathrm{ave} \equiv M_\mathrm{ave}^{(1)} + M_\mathrm{ave}^{(2)}$. The resulting $Q(H)$ curve fits well with the $Q^{(1)}$ and $Q^{(2)}$ curves obtained from the single-band analysis in Eqs.~\eqref{eq:Hnormalize} and \eqref{eq:Qnormalize}. The ground state in low magnetic field is identified as the $Q_2$-FFLO state with $n_2=0$ whose $Q(H)$ follows the $Q^{(2)}$ curve, while it turns to the $Q_1$-FFLO state with $Q(H)\sim Q^{(1)}(H)$ in the higher field regime. Figure~\ref{fig:Q_M-h_T-case1} clearly indicates that the transition from $Q_2$- to $Q_1$-FFLO phases is of the first-order. The transition between BCS to $Q_2$-FFLO state is also of the first-order, as shown in Fig.~\ref{fig:Q_M-h_T-case1}. By lowering the magnetic field, the transition occurs and $Q$ drops to zero around $\mu _\mathrm{B}H\approx 0.588\Delta^{(1)}_0$.

As shown in Fig.~\ref{fig:Q_M-h_T-case1}, the net magnetization jumps at each critical field. These jumps manifest the onset of the $Q_2$- and $Q_1$-FFLO phases and might be observable in experiments. In Fig.~\ref{fig:Q_M-h_T-case1}, it is also found that the net magnetization in the second band decreases at the critical field $\mu _\mathrm{B}H \approx 0.75\Delta^{(1)}_0$ between $Q_2$- and $Q_1$-FFLO phases, while the total magnetization increases. This is also observed in the case of $\mathcal{R} < 1$.

\begin{figure}[b]
\includegraphics[width=\linewidth]{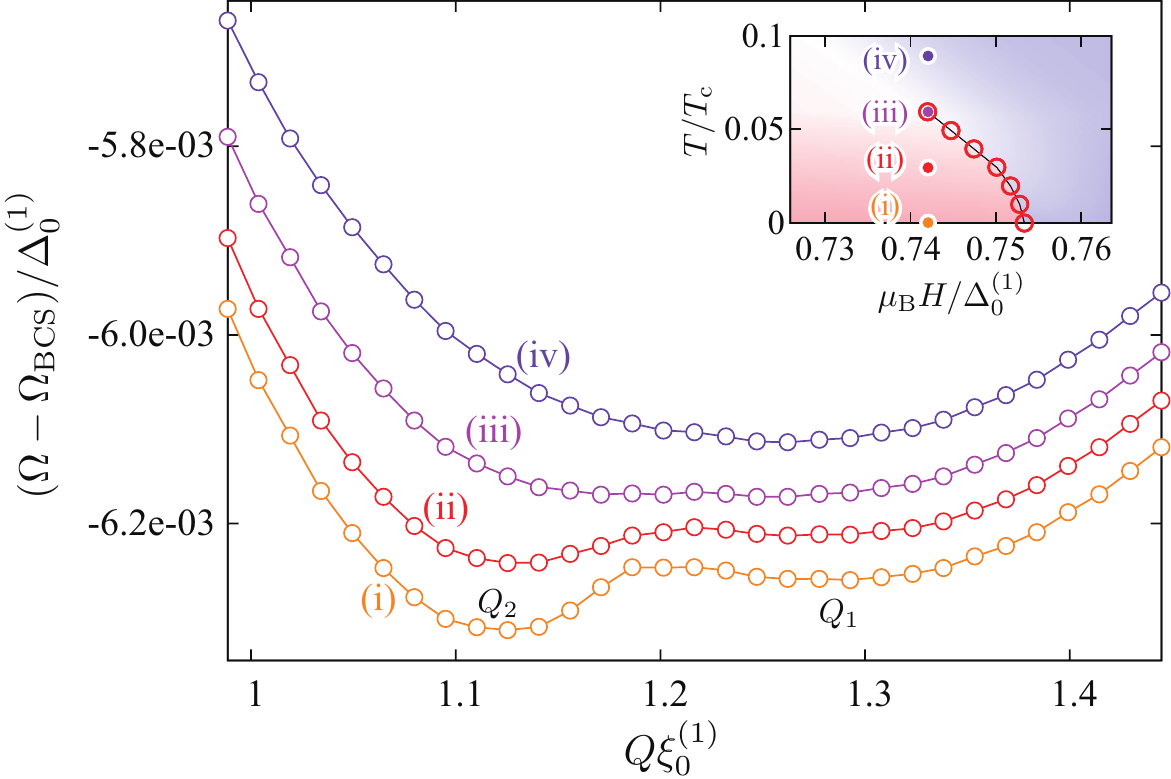}
\caption{(color online) Thermodynamic potential $\Omega$ as a function of the FFLO modulation wavenumber $Q$ at $\mu_\mathrm{B} H/\Delta_0^{(1)} = 0.74$ for various temperatures, $T/T_\mathrm{c} = 0$ (i), $0.03$ (ii), $0.06$ (iii), and $0.09$ (iv). Each plot is scaled by $i \times 5 \times 10^{-5} ~(i = 0, 1, 2, 3)$ for visibility. The inset shows the calculated points in the phase diagram in Fig.~\ref{fig:phase_T-case1}.}
\label{fig:Omega-Q_T-case1_ambiguos}
\end{figure}

As seen in Fig.~\ref{fig:phase_T-case1}, the phase boundary between the $Q_1$- and $Q_2$-FFLO states are indistinguishable in the high temperature region of the phase diagram. This is because the two local minima originated from ${Q}^{(1)}$ and ${Q}^{(2)}$ merge into the same value with increasing the temperature, as shown in Fig.~\ref{fig:Omega-Q_T-case1_ambiguos}. The $Q_2$-FFLO state with $n_2 = 0$ can be continuously connected to the $Q_1$-FFLO state with $n_1 = 0$ if they share the same $Q$. Note here that, as discussed later, this is different from the case of $\mathcal{R} < 1$ in which the family of the $Q_2$-FFLO states with $n_2 > 0$ becomes the ground state and successive first-order transitions occur.

\subsubsection{Energy spectra and density of states}

\begin{figure}[b]
\includegraphics[width=\linewidth]{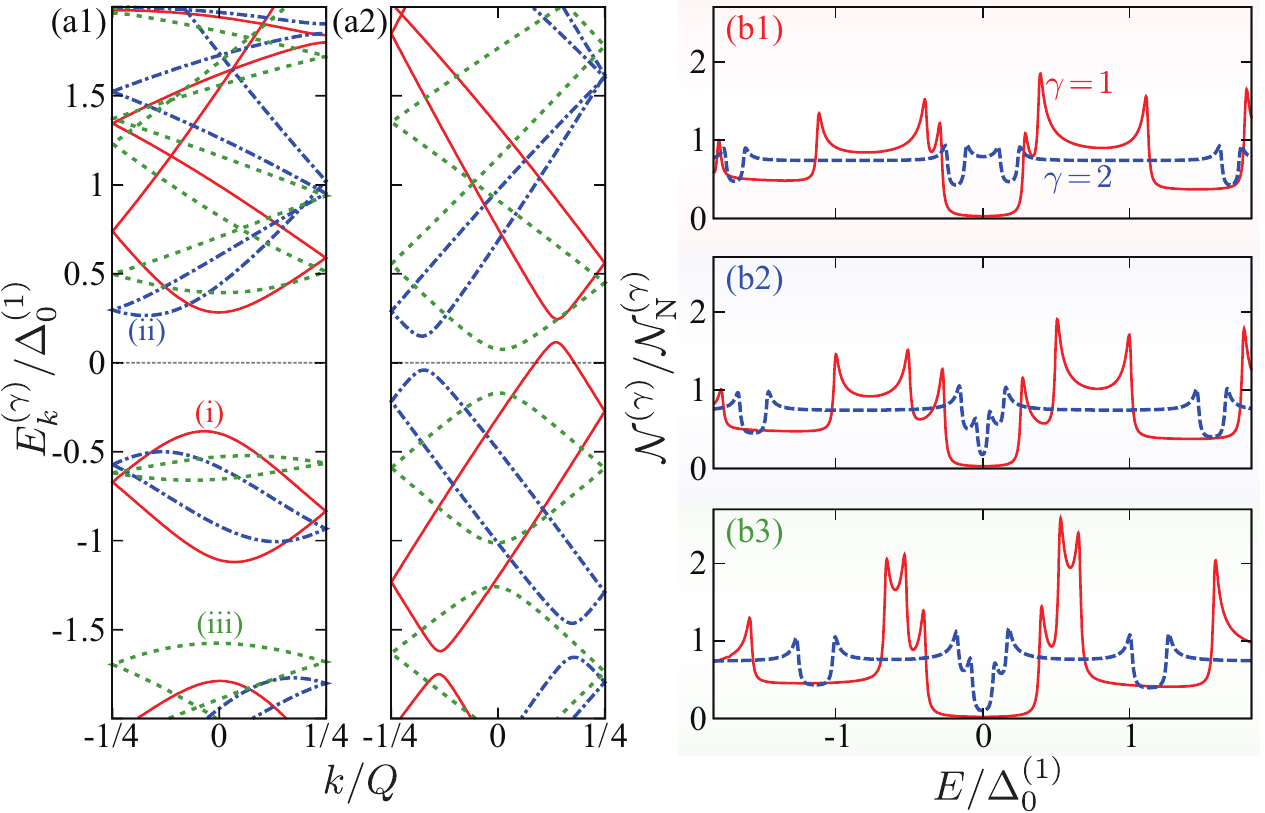}
\caption{(color online) Energy spectra of the first band (a1) and the second band (a2). The (red) solid line (i) corresponds to $Q_1$-FFLO states at $\mu_\mathrm{B} H / \Delta_0^{(1)} = 0.753$. The (blue) dash-dotted line (ii) and the (green) dotted line (iii) correspond to $Q_2$-FFLO states with $n_2 = 0$ at $\mu_\mathrm{B} H / \Delta_0^{(1)} = 0.753$ and $0.591$, respectively. The density of states in (i)-(iii) are displayed in (b1)-(b3), respectively. The other parameters are same as those in Fig.~\ref{fig:phase_T-case1}. 
}
\label{fig:spectra_DOS_T-case1}
\end{figure}

Let us confirm the realization of the $Q_2$-FFLO states from the perspective of the energy spectra. 
Figure~\ref{fig:spectra_DOS_T-case1} illustrates the energy spectra and the density of states (DOS) of three FFLO states. The corresponding pair potentials are displayed in Fig.~\ref{fig:delta_mag_T-case1}. The energy spectra in the first and second bands are plotted in Figs.~\ref{fig:spectra_DOS_T-case1}(a1) and \ref{fig:spectra_DOS_T-case1}(a2), respectively. The (red) solid line corresponds to the $Q_1$-FFLO state and the other lines depict the $Q_2$-FFLO states for various $H$'s. The energy spectrum of the first-band is qualitatively unchanged by the magnetic field, which always opens a finite energy gap at the Fermi level. 

The band structures around $E_k = - \mu_\mathrm{B} H$ in Figs.~\ref{fig:spectra_DOS_T-case1}(a1) and \ref{fig:spectra_DOS_T-case1}(a2) are interpreted as the lattice formation of the Jackiw-Rebbi solitons bound at the FFLO nodes.~\cite{jackiw} 
In the case of the $Q_2$-FFLO with $n_2=0$, as shown in Fig.~\ref{fig:spectra_DOS_T-case1}(a2), the energy dispersion of the soliton band does not cross the Fermi level. As $H$ increases, however, the FFLO modulation period $L = 2\pi/Q$
shortens and thus the dispersion of the Jackiw-Rebbi soliton band becomes dispersive. As shown in Fig.~\ref{fig:Q_M-h_T-case1}(a), $Q$ jumps at the critical $\mu _\mathrm{B}H/\Delta^{(1)}_0 \approx 0.75$ between the $Q_2$- and $Q_1$-FFLO phases. In the $Q_1$-FFLO phase, the Jackiw-Rebbi soliton band crosses the Fermi level. These features are also confirmed in the DOS of the second band (blue lines) displayed in Figs.~\ref{fig:spectra_DOS_T-case1}(b1)-(b3), where the zero-energy DOS is absent in the $Q_2$-FFLO states in Figs.~\ref{fig:spectra_DOS_T-case1}(b2) and \ref{fig:spectra_DOS_T-case1}(b3). The change of low-energy structures implies that the $Q_2$-FFLO states gain the condensation energy at low temperatures and thus they can be stable relative to the $Q_1$-FFLO state.

Since the soliton band is occupied (unoccupied) by only up-spin (down-spin) electrons in the low temperature regime, paramagnetic moments are accumulated by the FFLO nodal planes. Hence, the spatial oscillation of magnetization density in Figs.~\ref{fig:delta_mag_T-case1}(a2), \ref{fig:delta_mag_T-case1}(b2), and \ref{fig:delta_mag_T-case1}(c2), reflects the FFLO modulation period 
$L$ and Jackiw-Rebbi soliton states in the low temperatures, which might be detectable in experiments.

\subsection{$\mathcal{R} < 1$ case}
\label{subsec:R<1}

\subsubsection{Phase diagram}

\begin{figure}[t]
\includegraphics[width=\linewidth]{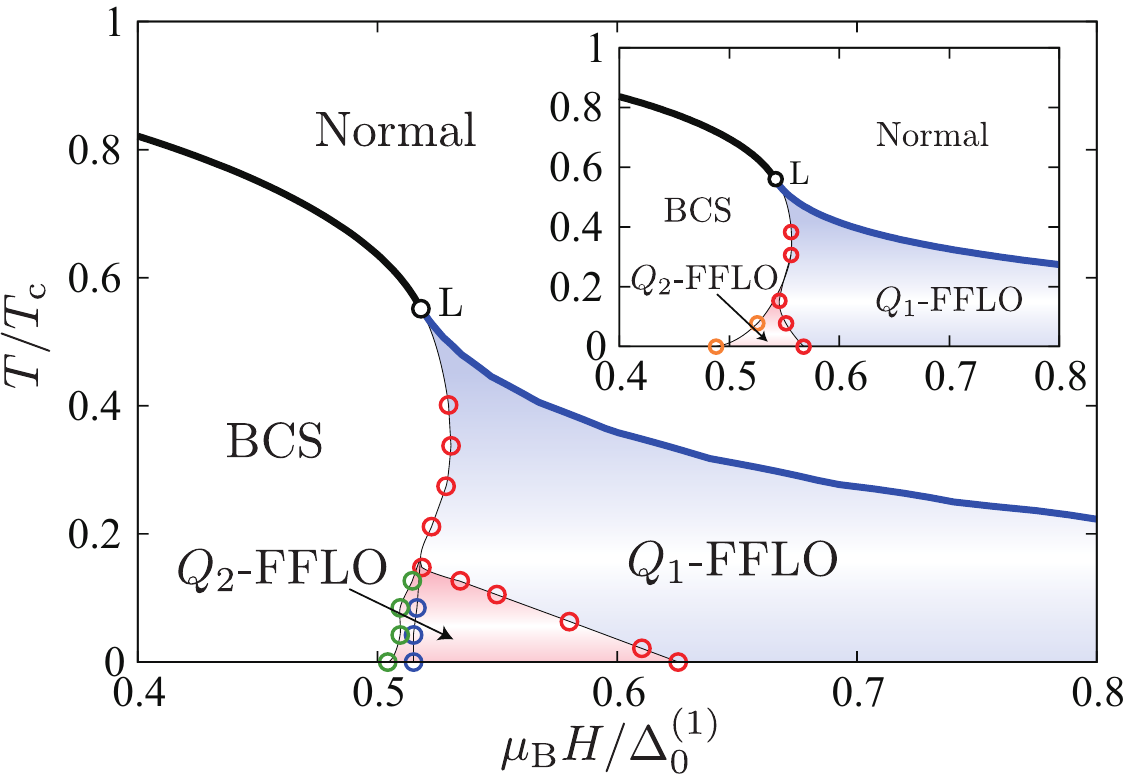}
\caption{(color online) Phase diagrams for $\bm{\Gamma} = (0.0625,0.5,0.2)$ in the main frame and for $\bm{\Gamma} = (0.1,0.6,0.5)$ in the inset. The former (latter) case corresponds to $\mathcal{R}=\sqrt{0.2}$ ($\sqrt{0.5}$) and $\Delta_0^{(2)}/\Delta_0^{(1)} \simeq 0.60$ ($0.53$). The FFLO phase is separated by the first-order transition to $Q_1$- and $Q_2$-FFLO states. The $Q_2$-FFLO phase is divided into a family of the $Q_2$-FFLO states through the first-order transitions.} 
\label{fig:phase_T-case2}
\end{figure}

Let us now turn to the case of $\mathcal{R} < 1$ where $Q^{(2)}$ is always bigger than $Q^{(1)}$. The resulting phase diagram is shown in Fig.~\ref{fig:phase_T-case2}. Here, we take the two sets of parameters: $\bm{\Gamma} = (0.0625,0.5,0.2)$ and $\bm{\Gamma} = (0.1, 0.6, 0.5)$. Using the former (latter) set, the ratio of the pair potential at $T=0$ and $H=0$ is estimated as $\Delta_0^{(2)} / \Delta_0^{(1)} \simeq 0.60$ ($0.53$). Then the ratio of the wavenumbers is $\mathcal{R} = \sqrt{0.2}$ and $\sqrt{0.5}$, respectively. For both the sets of parameters, the region of high fields and low temperatures is occupied by two FFLO phases, the $Q_1$- and $Q_2$-FFLO phases. In contrast to the case of $\mathcal{R}>1$, however, the $Q_2$-FFLO phase is further divided into a family of the $Q_2$-FFLO states through successive first-order phase transitions. The detailed structures will be discussed below. We here calculate the physical quantities in both the cases, which yield qualitatively same behaviors. In this paper, therefore, we focus on the results with the set of parameters, $\bm{\Gamma} = (0.0625,0.5,0.2)$. The more information of the latter case can be found in Ref.~\onlinecite{Mizushima_PRL}.

\begin{figure}[!t]
\includegraphics[width=\linewidth]{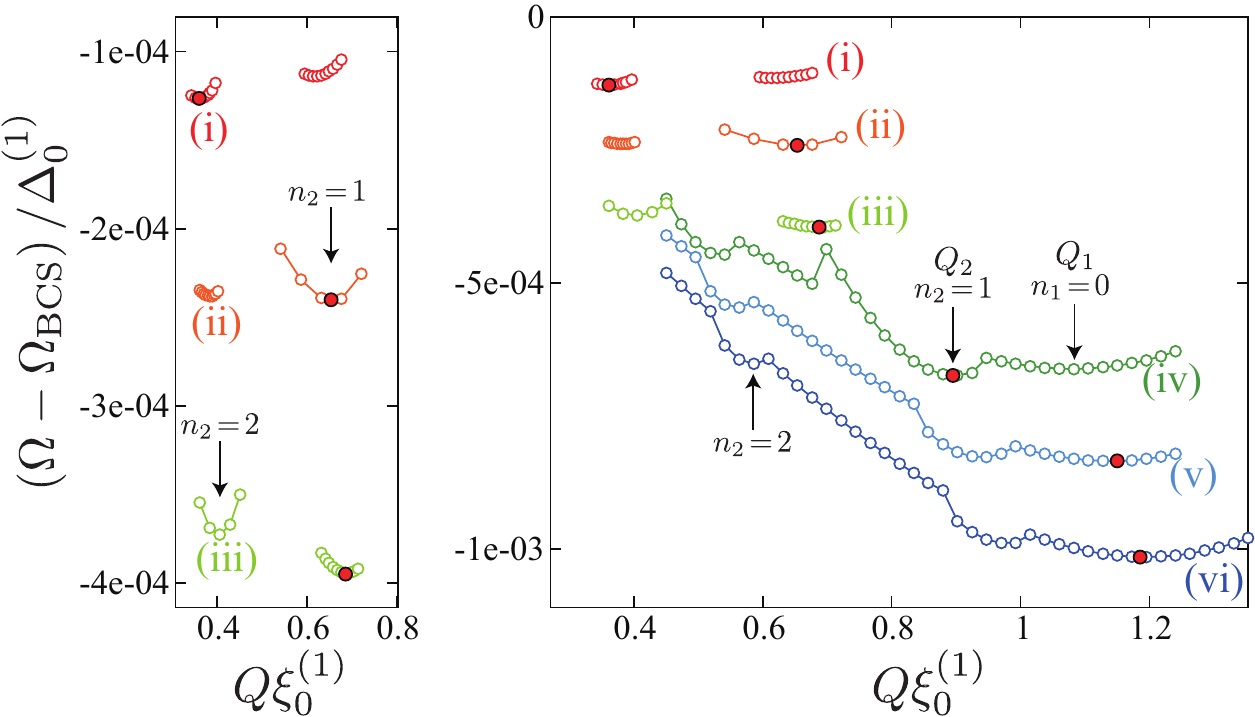}
\caption{(color online) Thermodynamic potential $\Omega(Q)$ at $T = 0$ for $\mu_\mathrm{B} H / \Delta_0^{(1)} =0.509$ (i), $0.516$ (ii), $0.525$ (iii), $0.605$ (iv), $0.625$ (v), and $0.646$ (vi). The set of parameters are $\bm{\Gamma} = (0.0625,0.5,0.2)$. The left panel is an enlarged plot of the low field region. The (red) filled circles denote the global minima of $\Omega(Q)$.}
\label{fig:Omega-Q_T-case2}
\end{figure}

\begin{figure*}[t!]
\begin{minipage}{0.75\linewidth}
\includegraphics[width=\linewidth]{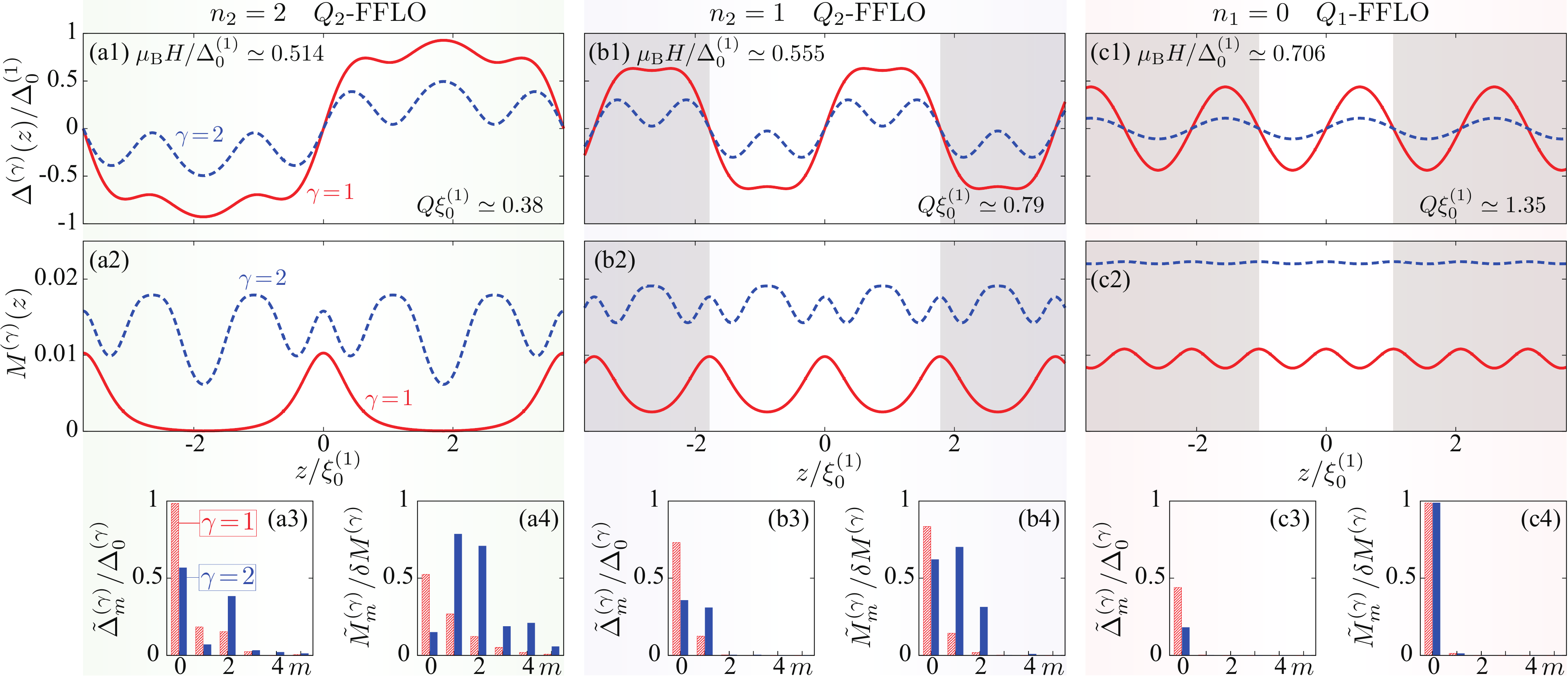}
\end{minipage}
\begin{minipage}{0.24\linewidth}
\caption{(color online) Spatial profiles of the gap for $\mu_\mathrm{B} H / \Delta_0^{(1)} = 0.514$ (a1), 
$0.555$ (b1), and $0.706$ (c1) at $T=0$. The corresponding magnetization densities are displayed in (a2), (b2), and (c2). (c1) and (c2) correspond to the $Q_1$-FFLO state.
(a1) and (a2) illustrate the $Q_2$-FFLO states with $n_2 = 2$ and (b1) and (b2) correspond to the $Q_2$-FFLO states with $n_2 = 1$. The parameters are same as which in Fig.~\ref{fig:phase_T-case2}. 
The bottom panels show their Fourier components where $\delta M^{(\gamma)} = \mathrm{max}[M^{(\gamma)}]-M_\mathrm{ave}^{(\gamma)}$. 
}
\label{fig:delta_mag_fourier_T-case2}
\end{minipage}
\end{figure*}

In Fig.~\ref{fig:Omega-Q_T-case2} we illustrate the thermodynamic potential $\Omega$ for various $H$'s as a function of the FFLO wavenumber $Q$. In Fig.~\ref{fig:Omega-Q_T-case2}, the data (i)-(iii) show $\Omega(Q)$ in the vicinity of the phase boundary between the BCS and $Q_2$-FFLO phases, and (iv)-(vi) are in the higher field regime around the boundary between $Q_2$- and $Q_1$-FFLO phases. There exist several local minima in $\Omega(Q)$. In the lower field regime, the FFLO phase with small $Q$'s can be the ground state, while the value of $Q$ in the ground state successively jumps to larger values with increasing $H$.

Figures~\ref{fig:delta_mag_fourier_T-case2}(a1)-(c1) show the spatial profile of the pair potentials 
and the corresponding magnetization densities are displayed in Figs.~\ref{fig:delta_mag_fourier_T-case2}(a2)-(c2).
The corresponding Fourier components are displayed in the bottom. 
As seen in Figs.~\ref{fig:delta_mag_fourier_T-case2}(c1) and~\ref{fig:delta_mag_fourier_T-case2}(c2), 
the pair potentials of two bands yield a simple sinusoidal modulation, and the spatial oscillations are characterized by a single FFLO modulation wavenumber, corresponding to the Fourier component with $m_{\gamma}=0$ in Eq.~\eqref{eq:Dm}. Similarly, the spatial modulation of $M^{(\gamma)}(z)$ is composed of a single Fourier component with $m=0$ in Eq.~\eqref{eq:Mm}. It turns out that the modulation wavenumber $Q$ coincides with the modulation wavenumber favored by the first band, $Q^{(1)}$. In this sense, we refer to this state as the $Q_1$-FFLO state. This state corresponds to the local-minimum state with the largest value of $Q$ in Fig.~\ref{fig:Omega-Q_T-case2}. The $Q_1$-FFLO state always occupies the high field region of the phase diagram, regardless of $\mathcal{R}>1$ and $\mathcal{R}<1$ (see Figs.~\ref{fig:phase_T-case1} and \ref{fig:phase_T-case2}) and the parameters $(g_{12}/g_{11},g_{22}/g_{11})$.

\subsubsection{Pair potentials and local magnetization}

Figures~\ref{fig:delta_mag_fourier_T-case2}(b1) and \ref{fig:delta_mag_fourier_T-case2}(a1) illustrate the spatial profiles of the pair potential which correspond to the local minima with the second and third largest $Q$ in Fig.~\ref{fig:Omega-Q_T-case2}, respectively. The corresponding magnetization densities are displayed in Fig.~\ref{fig:delta_mag_fourier_T-case2}(b2) and (a2). 
It is seen from Figs.~\ref{fig:delta_mag_fourier_T-case2}(b3) and \ref{fig:delta_mag_fourier_T-case2}(a3) that $\Delta^{(\gamma)}(z)$ is composed of multiple wavenumbers, 
$(2m+1)Q~(m = 1, 2, \dots)$, 
in addition to $Q$. In the case of Fig.~\ref{fig:delta_mag_fourier_T-case2}(b1), the wavenumber $3Q$ of $\Delta^{(\gamma)}(z)$ is favored by the pair potential of the second band $\Delta^{(2)}$, namely, $3Q \simeq Q^{(2)}$. Since the two bands are coherently coupled through the interband Cooper pair tunneling and $\Delta^{(1)}$ favors the relatively longer modulation period, the extra Fourier component with $Q$ then mixes the longer modulation period in $\Delta^{(\gamma)}(z)$, which is favored by the first band $\Delta^{(1)}$. As a result, the spatial modulation of Fig.~\ref{fig:delta_mag_fourier_T-case2}(b1) is described as $\Delta^{(\gamma)}(z) \approx \tilde{\Delta}^{(\gamma)}_0 \sin({Q^{(2)}z/3})+ \tilde{\Delta}^{(\gamma)}_1 \sin({Q^{(2)}z})$. We refer to this state as the $Q_2$-FFLO state with $n_2 =1$ (see Fig.~\ref{fig:Omega-Q_T-case2}). Similarly, $\Delta^{(\gamma)}(z)$ in the case of Fig.~\ref{fig:delta_mag_fourier_T-case2}(a1) is composed of mainly two Fourier components, $Q$ and $5Q$, where the shorter period $5Q \simeq Q^{(2)}$ is favored by the second band. The FFLO state in Fig.~\ref{fig:delta_mag_fourier_T-case2}(a1)
is identified as the $Q_2$-FFLO state with $n_2=2$. 

\begin{figure}[!t]
\includegraphics[width=\linewidth]{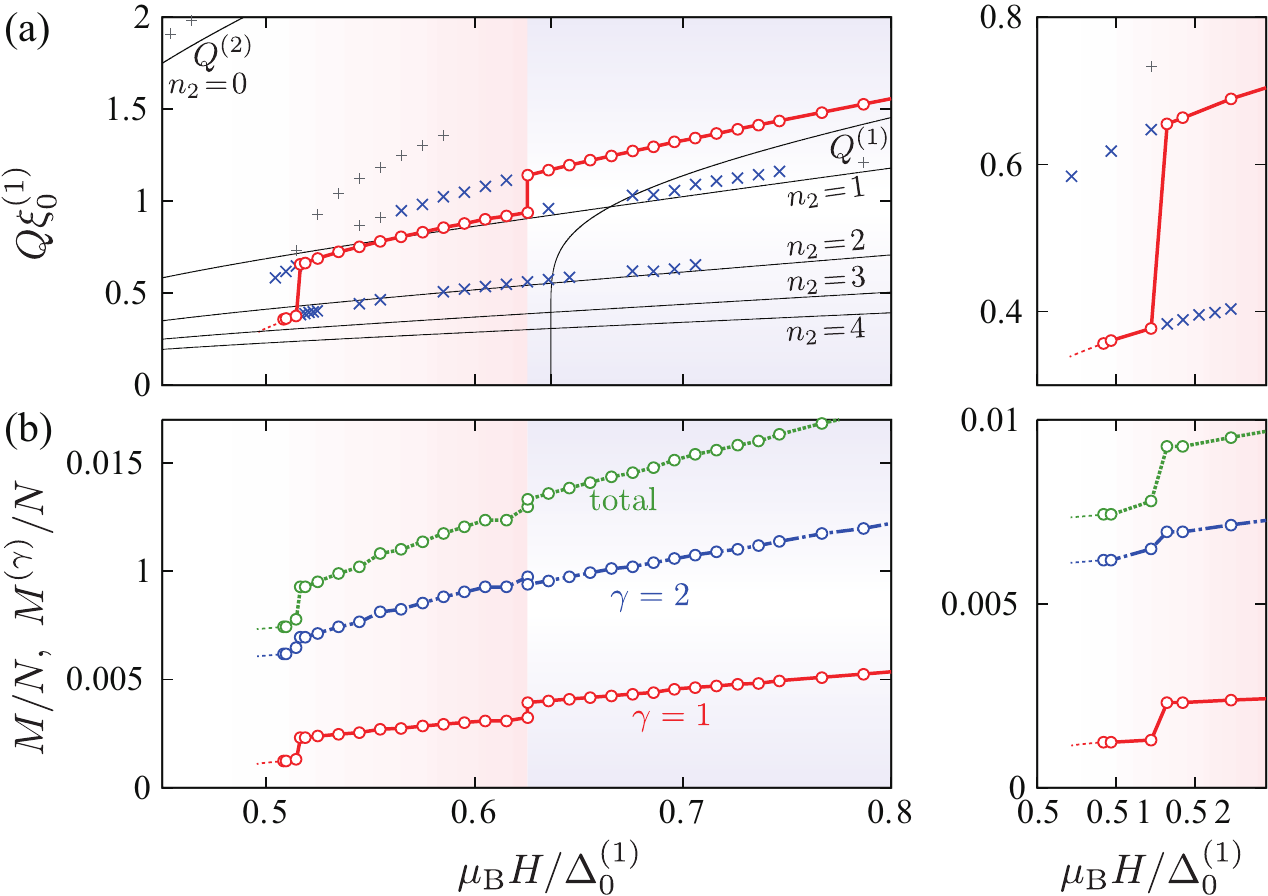}
\caption{(color online) (a) FFLO wavenumber $Q$ and (b)~spatially averaged magnetizations $M$ and $M^{(\gamma)}$ as a function of $H$ for $\bm{\Gamma} = (0.0625,0.5,0.2)$ and $T=0$. The (black) thin lines indicate the $Q(H)$ curve obtained form the single-band analysis. We also depict the curves $Q_2(H)/(2n_2+1)$, corresponding to the family of the $Q_2$-FFLO. The symbol``$\times$'' (``$+$'') denotes the local minimum state whose thermodynamic potential is lower (higher) than that in the BCS state. The right panels illustrate the detailed structures in the lower $H$ region. }
\label{fig:Q_M-h_T-case2}
\end{figure}

Note that in low temperatures, the modulation of the magnetization density reflects the spatial oscillation of $\Delta^{(\gamma)}(z)$ in the FFLO states. As seen in Fig.~\ref{fig:delta_mag_fourier_T-case2}(c4), the spatial modulation of the magnetization density in the $Q_2$-FFLO state is expressed by a single Fourier component with $2Q$. In Fig.~\ref{fig:delta_mag_fourier_T-case2}(b4) and (a4), we show the Fourier components of $M^{(\gamma)}(z)$ in $Q_2$-FFLO states with $n_2=2$ and $n_2=1$, where the spatial modulation consists of multiple Fourier components. This feature in $Q_2$-FFLO states is qualitatively different from the single-$Q$ behavior in the $Q_1$-FFLO state. Hence, the multiple-$Q$ modulations in the magnetization density manifest the stability of the $Q_2$-FFLO states.

Let us now consider the transition from $Q_2$- to $Q_1$-FFLO states. The field-dependence of the FFLO modulation wavenumber, $Q$, is displayed in Fig.~\ref{fig:Q_M-h_T-case2}(a). The corresponding phase diagram in Fig.~\ref{fig:phase_T-case2} indicates that in the higher magnetic field regime, the $Q_1$-FFLO state is a ground state. This is because the minority pair potential, $\Delta^{(2)}$, becomes passive in the high field regime and the FFLO modulation is determined by the majority band. Hence, the field-dependence of $Q$ follows the curve $Q^{(1)}(H)$ which is obtained from the single-band analysis in Eqs.~\eqref{eq:Hnormalize} and \eqref{eq:Qnormalize}. In the lower field regime where the pair potential in the second band is competitive to that of the first band, the frustration between $Q^{(1)}$ and $Q^{(2)}$ gives rise to anomalous field-dependence of the FFLO modulation wavenumber $Q$. In the case of $\mathcal{R} < 1$ which we consider, the FFLO wavenumber favored by the second-band, $Q^{(2)}$, is always larger than $Q^{(1)}$, as shown in Fig.~\ref{fig:analyticQs}(b). Hence, as discussed above, the lower field region is occupied by a family of the $Q_2$-FFLO states, whose pair potentials are characterized by multiple wavenumbers, 
$(2m+1)Q ~(m = 0, 1, \dots)$, 
with $Q \simeq Q^{(2)}/(2n_2+1)$. As illustrated in Fig.~\ref{fig:Q_M-h_T-case2}(a), the magnetic field drives successive first-order phase transition among $Q_2$-FFLO phases with different $n_2$'s, which are accompanied by the jump of the $Q$ value. The transition from $Q_2$- to $Q_1$-FFLO phases is also of the first-order. As seen in Fig.~\ref{fig:Q_M-h_T-case2}, the field-dependence of the net magnetization clearly indicates the successive first-order transitions. The so-called \textit{devil's staircase}~\cite{bak, chaikin} is peculiar to FFLO phases in multi-band superconductors and might be detectable in experiments.

Note that although there exist more steps in the $\mu _\mathrm{B}H/\Delta^{(1)}_0 \lesssim 0.5$ of Fig.~\ref{fig:Q_M-h_T-case2}(b), further calculations in the lower field region are technically difficult. This is because $Q_2$-FFLO states with an infinitely large $n_2$ are competing to the spatially uniform BCS state. We find that the thermodynamic potential of the BCS state is still higher than that of $Q_2$-FFLO states around $\mu _\mathrm{B}H/\Delta^{(1)}_0\sim 0.5$. Therefore, the critical field below which the BCS state is stabilized will be located in the region of $H^{(2)}_\mathrm{p} \lesssim H \lesssim 0.5\Delta^{(1)}_0/\mu _\mathrm{B}$, where $H^{(2)}_\mathrm{p}=\frac{1}{\sqrt{2}}\frac{\Delta^{(2)}_0}{\mu_\mathrm{B}}$ denotes the Pauli-limiting field in the second band and $\mu _\mathrm{B}H^{(2)}_\mathrm{p}/\Delta^{(1)}_0 \simeq
0.42$ with the parameters of Fig.~\ref{fig:Q_M-h_T-case2}.

\subsubsection{Energy spectra and density of states}

\begin{figure}[t]
\includegraphics[width=\linewidth]{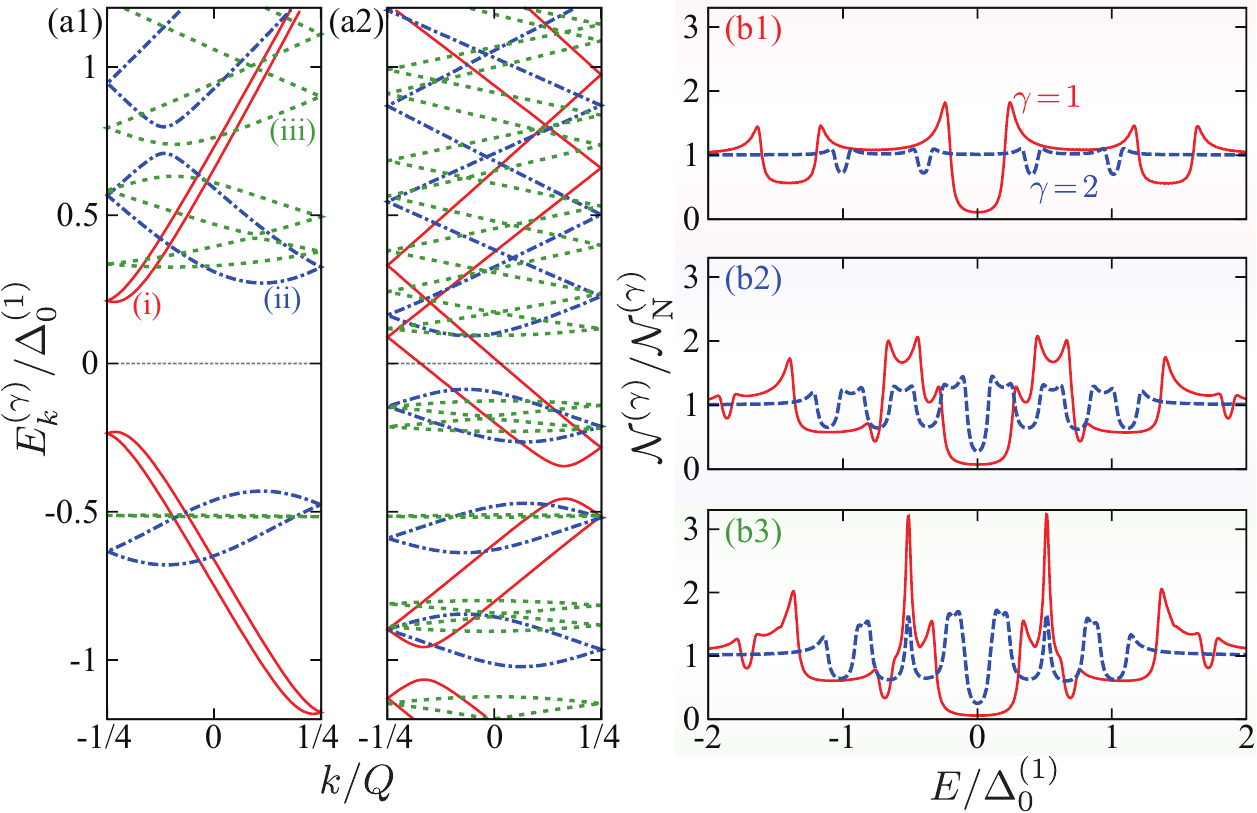}
\caption{(color online) Energy spectra of the first band (a1) and the second band (a2) for $\bm{\Gamma} = (0.0625,0.5,0.2)$. The (red) line (i) corresponds to $Q_1$-FFLO states at $\mu_\mathrm{B} H / \Delta_0^{(1)} = 0.706$. The (blue) dash-dotted line (ii) and (green) dotted line (iii) illustrate the $Q_2$-FFLO states with $n_2 = 1$ at $\mu_\mathrm{B} H / \Delta_0^{(1)} = 0.555$ and with $n_2 = 2$ at $0.514$, respectively. The corresponding density of states to the cases (i)-(iii) are displayed in (b1)-(b3). 
}
\label{fig:spectra_DOS_T-case2}
\end{figure}

Finally, to explain the thermodynamic stability of $Q_2$-FFLO states relative to the $Q_1$-FFLO state, we illustrate in Fig.~\ref{fig:spectra_DOS_T-case2} the quasiparticle energy spectra and DOS of the $Q_1$-FFLO state and $Q_2$-FFLO states with $n_2 = 1$ and $n_2=2$. Similarly with the case of $\mathcal{R}>1$ in Sec.~\ref{subsec:R>1}, the dispersion of the $Q_1$-FFLO state crosses the Fermi level, while the $Q_2$-FFLO states open a finite energy gap around the Fermi level. 

In the case of the $n_2=2$ $Q_2$-FFLO state, as shown in Fig.~\ref{fig:delta_mag_fourier_T-case2}(a3), $\Delta^{(\gamma)}(z)$ is composed of two modulation periods $2\pi /Q$ and $2\pi / (5Q)$, $\Delta^{(\gamma)}(z) \approx \tilde{\Delta}^{(\gamma)}_{0}\sin (Qz) + \tilde{\Delta}^{(\gamma)}_{2} \sin(5Qz)$. The FFLO state with the short period $2\pi/(5Q)$ is accompanied by a finite zero energy density of states.~\cite{machida1984} The mixing of the Fourier component with the longer modulation period $2\pi/Q$, however, folds back the original band into the small reduced Brillouin zone with the $1/5$ size. As a result, the band gap opens around the Fermi level in the reduced Brillouin zone of the minor band (Fig.~\ref{fig:spectra_DOS_T-case2}(a2)), resulting in the gain of condensation energy. The multiple-band structure appears around $E \sim -0.2\Delta^{(1)}_0$, ${\mu _\mathrm{B}H}$, and $-\Delta^{(1)}_0$ in the $Q_2$-FFLO states with $n_2 = 1$ and $n_2=2$. This helps the stability of the $Q_2$-FFLO states in the lower field regime, compared with the $Q_1$-FFLO state with a finite zero energy DOS.

\section{Conclusions}
\label{sec:summary}

We have studied multi-band effects on superconductivity under an external magnetic field. We have here focused on Fulde-Ferrell-Larkin-Ovchinnikov (FFLO) states in two-band systems. Extending the basic theory based on the Bogoliubov-de Gennes equation to a multi-band system, we have developed the formalism to calculate analytic phase boundaries between superconducting states and the normal state. 

We have first examined the phase diagram in a two-band system with a passive band, where intraband interaction in the second band is absent. As the interband interaction increases, the critical field, $H_\mathrm{LO}(T)$, above which the FFLO state is stable, decreases and becomes lower than the tricritical Lifshitz point. This indicates that the BCS-FFLO phase boundary line $T_\mathrm{LO}(H)$ yields a positive slope, $dT_\mathrm{LO}(H)/dH>0$, contrast to that in a single-band superconductor, $dT_\mathrm{LO}(H)/dH<0$.~\cite{machida1984} We have numerically checked that the temperature at the Lifshitz point is invariant with respect to interband and intraband interactions.

For a two-band superconductor where intraband interactions are present in both bands, there are two different FFLO modulation wavenumbers, $Q^{(1)}$ and $Q^{(2)}$, which are favored by the first and second bands, respectively. The competing effect between two length scales 
$2\pi/Q^{(1)}$
and $2\pi/Q^{(2)}$ 
plays an important role in the FFLO states of multi-band superconductors, giving rise to the existence of two FFLO phases: $Q_1$-FFLO and $Q_2$-FFLO phases. In this paper, we divide the parameter region into two cases: $\mathcal{R} \equiv \lim_{H \rightarrow H_{c2} (0)} Q^{(1)}/Q^{(2)} > 1$ and $\mathcal{R}<1$. In both cases, it has been demonstrated that the phase diagram in the low temperature and high field region is occupied by the $Q_1$-FFLO phase, whereas the $Q_2$-FFLO becomes stable in the lower field regime around the Pauli-limiting fields. In the case of $\mathcal{R} > 1$, the spatial modulation of the pair potential in the $Q_1$ and $Q_2$-FFLO phases is well characterized by a single modulation wavenumber $Q
\simeq 
Q^{(1)}(H)$ and $Q \simeq Q^{(2)}(H)$, respectively. The transition from the BCS to $Q_2$-FFLO phase is of the first order. With increasing an applied field, the $Q_2$-FFLO phase undergoes the first-order transition to the $Q_1$-FFLO phase in the low temperature regime, while the transition becomes ambiguous in high temperatures. In the case of $\mathcal{R}<1$, the $Q_2$-FFLO phase is divided into a family of the $Q_2$-FFLO states, where the FFLO modulation is composed of multiple wavenumbers. The transitions between all the superconducting phases are of the first order. The field-dependence of the FFLO modulation wavenumber and net magnetization is accompanied by the devil's staircase, which clearly indicates the successive first-order transitions between the $Q_2$-FFLO substates. 

Recently, strong Pauli-limited superconductivity has been reported in an Iron-based superconductor, KFe$_2$As$_2$.~\cite{burger,zocco} It is known that KFe$_2$As$_2$ has $\alpha$-, $\beta$-, $\zeta$-, and $\epsilon$-bands, where the $\epsilon$-band has a relatively larger Fermi velocity than that in other bands and $\Delta^{(\epsilon)}_0> \Delta^{(\alpha)}_0\sim\Delta^{(\beta)}_0\sim\Delta^{(\zeta)}_0$.~\cite{hardy} The FFLO modulation wavenumbers in each band are estimated as $Q^{(\alpha)}\sim Q^{(\beta)} \sim Q^{(\zeta)} < Q^{(\epsilon)}$.~\cite{Mizushima_PRL} This corresponds to the case of $\mathcal{R} >1$ in the present context. Hence, it is inferred that the $Q_2$-FFLO phase appears in the low temperature and high field regime and with increasing the magnetic field, it undergoes the first-order transition to the $Q_1$-FFLO phase. Note that the generalized WHH approach recently makes the Pauli-limit effect of KFe$_2$As$_2$ questionable.~\cite{koganRPP} To fully understand the phase diagram of KFe$_2$As$_2$, however, we must take account of orbital depairing effect which induces vortices. In a single-band superconductor, the formation of vortices is competitive to the stability of the FFLO state.~\cite{nakai,ichioka2004,ichioka2007,suzuki,TM2005v2,ichioka2007v2,suzuki2011} The study on the interplay of vortices with $Q_1$- and $Q_2$-FFLO states remains in the future work. ~

\acknowledgments

This work was supported by JSPS (Nos. 21340103, 23840034, 25287085, and 25800199) and by Grant-in-Aid for Scientific Research from MEXT of Japan, ``Topological Quantum Phenomena'' Nos.~22103005 and 25103716.


\appendix

\section{BdG equation}
\label{sec:bdgformalism}

In this Appendix, we describe the details on the derivation of the BdG equation \eqref{eq:BdGfinal} and gap equation \eqref{eq:gapfinal}.
Let us start from the second-quantized microscopic Hamiltonian with multiple bands under a uniform external magnetic field, which is presented in Eqs.~\eqref{eq:Hamiltonian}-\eqref{eq:int_part}.
Employing the mean field approximation, the Hamiltonian reduces to 
\begin{equation}
\mathcal{H}_\mathrm{int} \simeq E_\mathrm{cond} + \sum_\gamma \int d\bm{r} \left [ \Delta^{(\gamma) \ast} (\bm{r}) \psi_{\downarrow}^{(\gamma)} (\bm{r}) \psi_{\uparrow}^{(\gamma)} (\bm{r}) + \mathrm{h.c.} \right ]
\label{eq:Hmf}
\end{equation}
where
$E_\mathrm{cond} = -\sum_{\gamma, \gamma'} g_{\gamma \gamma'}^{-1} \int d\bm{r} \Delta^{(\gamma')\ast} (\bm{r}) \Delta^{(\gamma)} (\bm{r})$. 
Here, $\Delta^{(\gamma)}$ denotes the pair potential defined by
\begin{equation}
\Delta^{(\gamma)} (\bm{r}) \equiv \sum_{\gamma'} g_{\gamma \gamma'} \left \langle \psi_{\downarrow}^{(\gamma')} (\bm{r}) \psi_{\uparrow}^{(\gamma')} (\bm{r}) \right \rangle.
\label{eq:gapequation1}
\end{equation}

From now on, without loss of generality, $\bm{H}$ is fixed to the $z$ axis: $\bm{H} = (0,0,H)$. The second term of Eq.~\eqref{eq:HamiltonianDensity} becomes $-\mu_\mathrm{B}^{(\gamma)} H \underline{\sigma}^z$, and the single-particle Hamiltonian density~\eqref{eq:HamiltonianDensity} becomes diagonal.

It is convenient to introduce the following spinor notation:
$\bm{\Psi}^{(\gamma)} (\bm{r}) = [
\psi_{\uparrow}^{(\gamma)} (\bm{r}), \psi_{\downarrow}^{(\gamma)\dagger} (\bm{r}) ]^\mathrm{T}
$,
where $\mathrm{T}$ means transposition.
The mean-field Hamiltonian in Eq.~\eqref{eq:Hmf} is rewritten as follows:
\begin{equation}
\mathcal{H}_\mathrm{MF} = E^{\prime}_\mathrm{cond} + \sum_\gamma \int d\bm{r} \bm{\Psi}^{(\gamma)\dagger} (\bm{r}) \underline{\mathcal{K}}^{(\gamma)} (\bm{r}) \bm{\Psi}^{(\gamma)} (\bm{r}), 
\label{eq:meanfiledHamiltonian} 
\end{equation}
where the BdG Hamiltonian density is given by
\begin{equation}
\underline{\mathcal{K}}^{(\gamma)} (\bm{r}) 
= \begin{bmatrix}
\varepsilon^{(\gamma)} (\bm{r}) - \mu_\mathrm{B}^{(\gamma)} H & \Delta^{(\gamma)} (\bm{r}) \\
\Delta^{(\gamma)\ast} (\bm{r}) & -\varepsilon^{(\gamma)} (\bm{r}) - \mu_\mathrm{B}^{(\gamma)} H \end{bmatrix}. 
\label{eq:meanfieldHamiltonian}
\end{equation}

We introduce the following Bogoliubov transformation, 
\begin{align}
\bm{\Psi}^{(\gamma)} (\bm{r}) 
&= \sum_\nu \underline{U}_\nu^{(\gamma)} (\bm{r}) \bm{\eta}_\nu^{(\gamma)} \nonumber \\
&= \sum_\nu \begin{bmatrix}
u_\nu^{(\gamma)} (\bm{r}) & - v_\nu^{(\gamma)\ast} (\bm{r}) \\
v_\nu^{(\gamma)} (\bm{r}) & u_\nu^{(\gamma)\ast} (\bm{r}) \end{bmatrix} \bm{\eta}_\nu^{(\gamma)},
\label{eq:BogoliubovTransform}
\end{align}
with quasiparticle basis, 
$\bm{\eta}_\nu^{(\gamma)} = [\eta_{\uparrow,\nu}^{(\gamma)},  \eta_{\downarrow,\nu}^{(\gamma)\dagger}]^\mathrm{T}$.
With the transformation, the mean-field Hamiltonian \eqref{eq:meanfieldHamiltonian} with the BdG matrix is diagonalized to
\begin{equation}
\mathcal{H}_\mathrm{MF} = E^{\prime}_\mathrm{cond} + \sum _{\sigma,\nu,\gamma} E_{\sigma,\nu}^{(\gamma)} \eta_{\sigma,\nu}^{(\gamma)\dagger} \eta_{\sigma,\nu}^{(\gamma)}. 
\label{eq:diag}
\end{equation}
The creation and the annihilation operators of the quasiparticles, $\eta_{\sigma,\nu}^{(\gamma)\dagger}$ and $\eta_{\sigma,\nu}^{(\gamma)}$, satisfy fermionic anti-commutation relations. The quasiparticle wave function in the matrix form $\underline{U}_\nu^{(\gamma)} (\bm{r})$ satisfies the orthonormal condition,
\begin{equation}
\int d\bm{r} \underline{U}_\nu^{(\gamma)\dagger} (\bm{r}) \underline{U}_{\nu'}^{(\gamma)} (\bm{r}) = \delta_{\nu,\nu'} \underline{1}_2,
\label{eq:orthonormal}
\end{equation}
and the completeness condition,
\begin{equation}
\sum_\nu \underline{U}_\nu^{(\gamma)} (\bm{r}) \underline{U}_\nu^{(\gamma)\dagger} (\bm{r}') = \delta (\bm{r} - \bm{r}') \underline{1}_2.
\label{eq:completeness}
\end{equation}
By substituting Eq.~\eqref{eq:BogoliubovTransform} to Eq.~\eqref{eq:meanfiledHamiltonian} and using Eqs.~\eqref{eq:diag} and \eqref{eq:completeness}, we derive the following Bogoliubov-de Gennes (BdG) equation
\begin{equation}
\underline{\mathcal{K}}^{(\gamma)} (\bm{r}) \underline{U}_\nu^{(\gamma)} (\bm{r}) = \underline{U}_\nu^{(\gamma)} (\bm{r}) \begin{bmatrix} E_{\uparrow,\nu}^{(\gamma)} & 0 \\
0 & - E_{\downarrow,\nu}^{(\gamma)} \end{bmatrix}.
\label{eq:BdGequation}
\end{equation}
The BdG matrix $\underline{\mathcal{K}}^{(\gamma)} (\bm{r})$ described above yields double eigenstates for spins. To see this, we set the eigenfunction of up spin states with an eigenvalue $E_{\uparrow,\nu}^{(\gamma)}$ as $\bm{\varphi}_{\uparrow,\nu}^{(\gamma)} (\bm{r}) = [ u_\nu^{(\gamma)} (\bm{r}), v_\nu^{(\gamma)} (\bm{r}) ]^\mathrm{T}$. Then, the corresponding BdG equation is given by $\underline{\mathcal{K}}^{(\gamma)} (\bm{r}) \bm{\varphi}_{\uparrow,\nu}^{(\gamma)} (\bm{r}) = E_{\uparrow,\nu}^{(\gamma)} \bm{\varphi}_{\uparrow,\nu}^{(\gamma)} (\bm{r})$. It is found that the BdG equation \eqref{eq:BdGequation} simultaneously has eigenstates for down spins of $\bm{\varphi}_{\downarrow,\nu}^{(\gamma)} (\bm{r}) = [ - v_\nu^{(\gamma)\ast} (\bm{r}), u_\nu^{(\gamma)\ast} (\bm{r}) ]^\mathrm{T}$ with the eigenvalue of~$- E_{\downarrow,\nu}^{(\gamma)}$: $\underline{\mathcal{K}}^{(\gamma)} (\bm{r}) \bm{\varphi}_{\downarrow,\nu}^{(\gamma)} (\bm{r}) = -E_{\downarrow,\nu}^{(\gamma)} \bm{\varphi}_{\downarrow,\nu}^{(\gamma)} (\bm{r})$. Therefore, once one solves the BdG equation for the quasiparticle $\bm{\varphi}_\nu^{(\gamma)} (\bm{r}) = [u_\nu^{(\gamma)} (\bm{r}), v_\nu^{(\gamma)} (\bm{r}) ]^\mathrm{T}$ and $E_\nu^{(\gamma)}$,
\begin{equation}
\underline{\mathcal{K}}^{(\gamma)} (\bm{r}) \bm{\varphi}_{\nu}^{(\gamma)} (\bm{r}) = E_\nu^{(\gamma)} \bm{\varphi}_{\nu}^{(\gamma)} (\bm{r}),
\label{eq:BdGfinalapp}
\end{equation}
then the solution gives the eigenstate for both spin states with positive and negative energies.

One can realize that the BdG equation \eqref{eq:BdGfinalapp} is the same form as that in a single band superconductor. Multi-band effects appear through the pair potential $\Delta^{(\gamma)}(\bm{r})$. The gap equation is derived by substituting Eq.~\eqref{eq:BogoliubovTransform} to Eq.~\eqref{eq:gapequation1} and replacing variables in Eq.~\eqref{eq:BdGfinalapp}, 
\begin{align}
\Delta^{(\gamma)} (\bm{r}) 
=& \sum _{\gamma^{\prime},\nu}g_{\gamma \gamma'} \Bigl [ u_\nu^{(\gamma')} (\bm{r}) v_\nu^{(\gamma')\ast} (\bm{r}) f (E_{\uparrow,\nu}^{(\gamma')}) \nonumber \\
& - v_\nu^{(\gamma')\ast} (\bm{r}) u_\nu^{(\gamma')} (\bm{r}) f(-E_{\downarrow,\nu}^{(\gamma')}) \Bigr ]\\
=& \sum _{\gamma^{\prime},\nu}g_{\gamma \gamma'} u_\nu^{(\gamma')} (\bm{r}) v_\nu^{(\gamma') \ast} (\bm{r}) f (E_\nu^{(\gamma')}), 
\label{eq:gapequation2}
\end{align}
where $f (E) = 1/(e^{\beta E} + 1)$ is the Fermi distribution function with inverse temperature $\beta = 1/(k_\mathrm{B} T)$ and $k_\mathrm{B}$ is the Boltzmann constant. Here, we utilize the following relations, 
$\langle \eta_{\sigma,\nu}^{(\gamma)} \eta_{\sigma',\nu'}^{(\gamma)} \rangle 
= \langle \eta_{\sigma,\nu}^{(\gamma)\dagger} \eta_{\sigma',\nu'}^{(\gamma)\dagger} \rangle = 0$,
$\langle \eta_{\uparrow,\nu}^{(\gamma)\dagger} \eta_{\uparrow,\nu'}^{(\gamma)} \rangle 
= f(E_{\uparrow,\nu}^{(\gamma)}) \delta_{\nu,\nu'}$, and 
$\langle \eta_{\downarrow,\nu}^{(\gamma)} \eta_{\downarrow,\nu'}^{(\gamma)\dagger} \rangle = f (-E_{\downarrow,\nu}^{(\gamma)}) \delta_{\nu,\nu'}$. 
The derived gap equation \eqref{eq:gapequation2} has a finite energy cutoff $E_\mathrm{c}$ in numerics, 
\begin{equation}
\Delta^{(\gamma)} (\bm{r}) = \sum_{\gamma'} \sum_{|E_\nu^{(\gamma')}| < E_\mathrm{c}} g_{\gamma \gamma'} u_\nu^{(\gamma')} (\bm{r}) v_\nu^{(\gamma') \ast} (\bm{r}) f (E_\nu^{(\gamma')}).
\label{eq:gapfinalapp}
\end{equation}
In the case of a single-band superconductor with a spatially uniform gap $\Delta_\mathrm{s}$, the gap equation reduces to 
\begin{equation}
\frac{1}{|g_{11}|} = \frac{1}{2\pi} \int_0^{k_\mathrm{c}} \frac{d k}{\sqrt{(k^2-\mu_1)^2 + \Delta_\mathrm{s}^2}}, 
\label{eq:g11}
\end{equation}
where $k_\mathrm{c} = \sqrt{2 M_\mathrm{e} E_\mathrm{c}/\hbar^2}$. 
This gives the relation between the cutoff energy and physical quantities such as the intraband coupling constant $g_{11}$ and the pair potential $\Delta_\mathrm{s}$ at $T = 0$ and $H = 0$. Hence, we estimate the amplitude of the intraband coupling constant $g_{11}$ from Eq.~\eqref{eq:g11} with the energy cutoff $E_\mathrm{c}$, the chemical potential $\mu _1$, and the pair potential $\Delta _\mathrm{s}$ at $T = 0$ and $H = 0$.

\section{Solving the BdG equation with a periodic boundary condition}
\label{sec:B}

\subsection{BdG equation with a periodic boundary condition}

We start by considering a periodic boundary condition, 
\begin{equation}
\Delta (\bm{r}+\bm{R}) = e^{i\chi(\bm{R})} \Delta (\bm{r}),
\label{eq:period}
\end{equation}
where $\chi (\bm{R}) \in \mathbb{R}$. The Bravais lattice vector is defined as $\bm{R} = n_1 \bm{a}_1 + n_2 \bm{a}_2 + n_3 \bm{a}_3$ with $n_j \in \mathbb{Z}$ and the primitive vectors of the lattice, $\bm{a}_j$ ($j=1,2,3$). We assume that the phase factor $\chi(\bm{R})$ yields additivity, $\chi (\bm{R}) + \chi(\bm{R}^{\prime}) = \chi(\bm{R} + \bm{R}^{\prime})$. For simplicity, we do not take account of the local gauge transformation, which implies that $\chi(\bm{R})$ is independent of $\bm{r}$.

The BdG equation is 
\begin{equation}
\mathcal{H}(\bm{r}) \bm{\varphi}_{\mu}(\bm{r}) = E_{\mu}\bm{\varphi}_{\mu}(\bm{r}), 
\label{eq:bdg_append}
\end{equation}
where the matrix is given by
\begin{equation}
\mathcal{H}(\bm{r}) = 
\begin{bmatrix}
\epsilon (\bm{r}) & \Delta (\bm{r}) \\ \Delta (\bm{r}) & - \epsilon^{\ast}(\bm{r})
\end{bmatrix}.
\label{eq:bdgmat}
\end{equation}
Let us now consider the periodic boundary condition for the BdG equation given by $T_{\bm{R}}[\mathcal{H}(\bm{r}) \bm{\varphi}_{\mu}(\bm{r})] = E_{\mu}T_{\bm{R}}\bm{\varphi}_{\mu}(\bm{r})$, where the translational operator, $T_{\bm{R}}$, obeys $T_{\bm{R}} T_{\bm{R}^{\prime}} = T_{\bm{R}^{\prime}}T_{\bm{R}} =T_{\bm{R}+\bm{R}^{\prime}}$. It is obvious from \eqref{eq:period} that the periodicity of the BdG matrix \eqref{eq:bdgmat} is expressed as the U(1) gauge transformation, $T_{\bm{R}}[\mathcal{H}(\bm{r})] = U(\bm{R})\mathcal{H}(\bm{r})U(\bm{R})$, where $U(\bm{R})= e^{i\chi(\bm{R})\sigma _z /2}$ yields additivity, $U(\bm{R}) + U(\bm{R}^{\prime}) = U(\bm{R}+\bm{R}^{\prime})$ and is supposed to be $[U(\bm{R})]^4=1$.  In addition, the eigenfunction of the BdG equation, $\bm{\varphi}_{\mu}$, is the simultaneous eigenfunction of the translation operator,
\begin{equation}
\left[
\mathcal{H}(\bm{r}), U(\bm{R})^{\dag}T_{\bm{R}}
\right] = 0.
\end{equation}
Using the Bloch vector $\bm{k} = \frac{m_1}{N_1} \bm{b}_1 + \frac{m_2}{N_2} \bm{b}_2 + \frac{m_3}{N_3} \bm{b}_3$ with the reciprocal lattice vectors $\bm{b}_i = 2\pi \epsilon _{ijk}(\bm{a}_j \times \bm{a}_k)/[\bm{a}_i\cdot(\bm{a}_j\times\bm{a}_k)]$, we obtain the periodic boundary condition on the quasiparticle wavefunction as
\begin{equation}
\bm{\varphi}_{\mu}(\bm{r}+\bm{R}) = U(\bm{R})e^{i\bm{k}\cdot\bm{R}}\bm{\varphi}_{\mu}(\bm{r}).
\label{eq:periodbc}
\end{equation}
We here impose the Born-von Karman boundary condition, $\bm{\varphi}_{\mu}(\bm{r}+N_j \bm{a}_j) = \bm{\varphi}_{\mu}(\bm{r})$, where $N_j \in \mathbb{Z}$ and $m_j \in \mathbb{Z}$.

In the current problem, we set $R=L/2$ in Eq.~\eqref{eq:period}. The phase shift is taken to be $\chi = \pi$ for FFLO states and $\chi = 2\pi$ for spatially uniform BCS states. Then, the periodic boundary condition on the quasiparticle wavefunction is recast into
\begin{equation}
\bm{\varphi}_{\mu}\left(z+\frac{L}{2}\right) 
= e^{i\bm{k}\cdot\bm{R}}e^{i\chi \sigma _z/2}\bm{\varphi}_{\mu}(z),
\label{eq:period2}
\end{equation}
where the Bloch vector is taken to be $\bm{k}= 2\pi m\hat{\bm{z}}/(LN_L)$ with $N_L\equiv N_{j}/2$ and $-N_L \le m < N_L$. We numerically solve the resulting one-dimensional BdG equation with the periodic boundary condition in Eq.~\eqref{eq:period2} using the finite-element method with the Gauss-Lobatto discrete variable representation

\subsection{Finite element method with the Gauss-Lobatto discrete variable representation}

Here we describe the details about how to numerically solve the BdG equation \eqref{eq:bdg_append} with the periodic boundary condition in Eq.~\eqref{eq:period2}. In order to map the BdG equation into the eigenvalue equation, we apply the the finite (FE) element method implemented with the Gauss-Lobatto discrete variable representation (DVR).~\cite{fedvr}

\subsubsection{Finite-element method with a periodic boundary condition}

For simplicity, let us consider the following differential equation in the coordinate space, $q \!\in\! [R_L,R_{R}]$,
\begin{equation}
H(q)\Psi(q) = E\Psi(q).
\end{equation}
Without loss of generality, $H(q)$ is composed of spatial derivatives and potential terms as $H(q) = a\frac{d^2}{dq^2} + V(q)$. The eigenfunction $\Psi(q)$ obeys the periodic boundary condition, 
\begin{equation}
\Psi(q=R_L) = \mathcal{S} \Psi(q=R_R),
\label{eq:phase_factor}
\end{equation}
where $\mathcal{S}\in\mathbb{C}$ and $|\mathcal{S}|=1$. 

\begin{figure}[t!]
\includegraphics[width=0.7\linewidth]{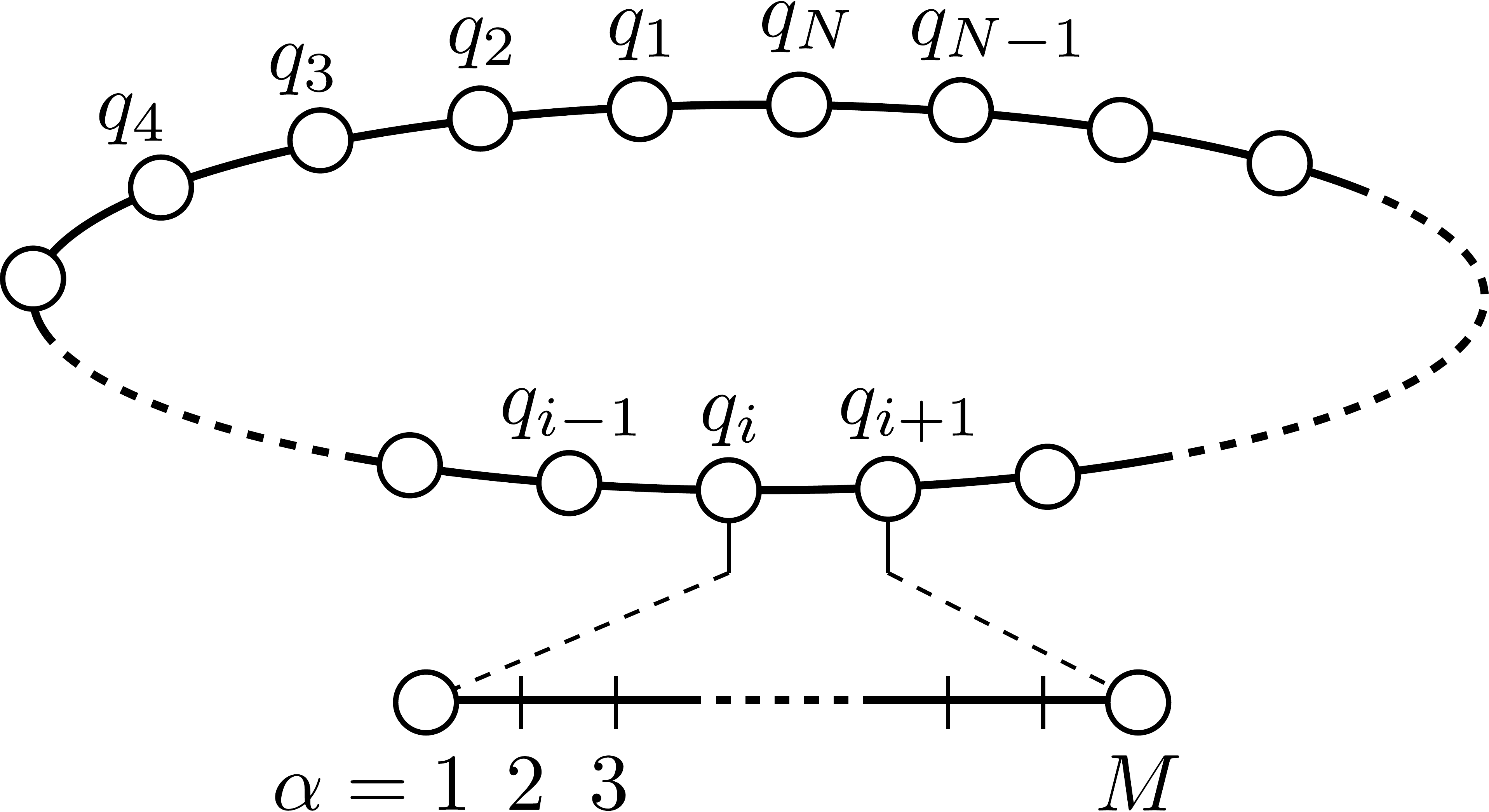}
\caption{Grids used in the FE-DVR method.} 
\label{fig:grids}
\end{figure}

We start by dividing the continuous variable $q$ into the $N$ intervals, where the $i$-th interval is defined as $q\in[q_i,q_{i+1}]$ with the $N+1$ discrete grids $\{q_i \}_{i=0, \dots, N}$ $(q_0 < q_1 < q_2 < \dots < q_{N}$). Each interval is further divided into sub grids labeled by $\alpha =1, \dots, M$ as $\{ q^{i}_{\alpha} \}$ (see Fig.~\ref{fig:grids}). In order to represent wave functions in each interval, we define the set of finite-element basis functions, $\{ f_{i,\alpha}(q)\}_{i=1, \dots, N}$, which are identically zero outside a given interval:  $f_{i,\alpha}(q)=0$ for $q\notin [q_i,q_{i+1}]$. The index $\alpha = 1, \dots, M$ denotes the local basis function in each interval and represents the DVR grids.~\cite{dvr1,dvr2,dvr3}
Following Ref.~\onlinecite{fedvr}, we introduce the boundary condition on 
\begin{equation}
f_{i,\alpha=1}(q_i)=1, \hspace{3mm} f_{i,\alpha=M}(q_{i+1})=1,
\end{equation}
and for $\alpha = 2,\dots, M-1$ 
\begin{equation}
f_{i,\alpha}(q_i) = f_{i,\alpha}(q_{i+1}) = 0 .
\end{equation}

It is convenient to construct the set of the orthonormal basis functions, $\{ \chi _{i,\alpha}(q) \}$, from $\{f_{i,\alpha}(q)\}$, which obeys the condition, 
\begin{equation}
\int^{R_R}_{R_L} \chi^{\ast}_{i,\alpha} (q) \chi _{j,\beta}(q) dq = \delta _{i,j}\delta _{\alpha,\beta}.
\label{eq:norm_dvr}
\end{equation}
We expand the eigenfunction $\Psi(q)$ in terms of the basis functions, 
\begin{equation}
\Psi(q) = \sum _{i,\alpha}C_{i,\alpha}\chi _{i,\alpha}(q).
\label{eq:psi}
\end{equation}
Since the eigenfunction must be continuous across each interval, the coefficients for $i=2, \dots, N-1$ are required to satisfy the following conditions
\begin{equation}
C_{i,M} = C_{i+1,1} ,
\label{eq:period1}
\end{equation}
for $i=2, \dots, N-1$ and 
\begin{equation}
C_{0,\alpha} = \mathcal{S} C_{N,\alpha}.
\end{equation}
This indicates that the coordinate $q\in [R_L,R_R]$ is divided into $(M-1)\times N + 1$ grid points. Each grid $q^{i}_{\alpha}$ is accompanied by the local basis functions $\chi _{i,\alpha}(q)$. 

\subsubsection{The Gauss-Lobatto discrete variable representation}

We now explicitly define the local basis functions, $\{f_{i,\alpha}(q)\}$, that is, $\{ \chi _{i,\alpha}\}$. Following Ref.~\onlinecite{fedvr}, the DVR is implemented to the finite-element method. The discrete grids in each interval, $\{q^{i}_{\alpha}\}$, are chosen to be a set of Gauss-Lobatto quadrature points. 

In the interval of $\bar{q}\in [-1,+1]$, the Gauss-Lobatto quadrature rule turns the integral of an arbitrary function $G(\bar{q})$ over the continuous variable $\bar{q}$ to the summation on the discrete grids $\bar{q}_{\alpha}$, $\int^{+1}_{-1} G(\bar{q}) d\bar{q} \!\approx\! G(-1)\bar{w}_1 + \sum^{M-1}_{\alpha=2} G(\bar{q}_{\alpha}) \bar{w}_{\alpha}+G(+1)\bar{w}_M$, which gives an exact result if the function $G(\bar{q})$ is a polynomial of degree $\le\! 2M-3$. In contrast to the Gauss-Legendre quadrature, the endpoints are constrained to $\bar{
q}_{\alpha = 1} = q_i$ and $\bar{q}_{\alpha=M}=q_{i+1}$. The points and weights $(\bar{q}_{\alpha},\bar{w}_{\alpha})$ are chosen to be the $\alpha$-th zeroth of the first derivative of the $(M-1)$-th degree Legendre polynomials, $dP_M(\bar{q})/\bar{q}$, and 
\begin{equation}
\bar{w}_{\alpha} = \left\{
\begin{array}{ll}
\displaystyle{\frac{2}{M(M-1)}}, & \mbox{for $\alpha = 1, M$}; \\
\\
\displaystyle{\frac{2}{M(M-1)[P_{M-1}(\bar{q}_{\alpha})]^2}}, & \mbox{otherwise}. 
\end{array} 
\right. 
\end{equation} 
Using the Gauss-Lobatto quadrature points and weights, we now define the grid points $q^{i}_{\alpha}$ and the weights $w^{i}_{\alpha}$ in each interval as
\begin{gather}
q^{i}_{\alpha} = \frac{(q_{i+1}-q_i)\bar{q}_{\alpha} + (q_{i+1}+q_i)}{2}, \\
w^{i}_{\alpha} = \frac{q_{i+1}-q_i}{2}\bar{w}_{\alpha}.
\end{gather}

Now let us construct the DVR basis functions based on the Gauss-Lobatto quadrature, where arbitrary functions given at each FE-DVR grids $\{q^{i}_{\alpha}\}$ are interpolated by Lagrange polynomials, 
\begin{equation}
f_{i,\alpha}(q) = \left\{
\begin{array}{ll}
\displaystyle{\prod _{\beta\neq \alpha}\frac{q-q^{i}_{\beta}}{q^{i}_{\alpha}-q^{i}_{\beta}}}, & 
 q_i \le q \le q_{i+1}; \\
\\
0, & q\notin[q_i,q_{i+1}].
\end{array}
\right. 
\end{equation}
The set of the basis functions is localized at each FE-DVR grid, 
\begin{equation}
f_{i,\alpha}(q^{j}_{\beta}) = \delta  _{i,j} \delta _{\alpha,\beta}, 
\end{equation}
and obeys the orthogonal condition with the accuracy of the Gauss-Lobatto quadrature, $\int^{R_R}_{R_L}f_{i,\alpha}(q)f_{j,\beta}(q)dq \approx w^{i}_{\alpha}\delta _{i,j}\delta _{\alpha,\beta}$. 

We can now construct the set of orthonormal functions, $\{ \chi _{i,\alpha}(q)\}$, from the local basis functions, $\{ f_{i,\alpha}(q)\}$, which allows to expand the eigenfunction $\Psi(q)$. The basis functions are continuous in each interval, $q\in [q_i,q_{i+1}]$, while the continuity at the interval boundaries is ensured by introducing the so-called ``bridge'' function, $\{\chi _{i,\alpha}\}$. The normalized bridge function at the grid $q^{i}_{\alpha}=q^{1}_{1}$ and $q^{N}_{M}$ is defined as
\begin{equation}
\chi _{i,\alpha}(q) = \frac{f_{N,M}(q) + f_{1,1}(q)}{\sqrt{w^{N}_{M}+w^{1}_{1}}}.
\label{eq:chi1}
\end{equation} 
The bridge functions for interval boundaries with $i\neq 1$ and $\alpha = M$ are given by 
\begin{equation}
\chi _{i,\alpha}(q) = \frac{f_{i,M}(q) + f_{i+1,1}(q)}{\sqrt{w^{i}_{M}+w^{i+1}_{1}}}.
\label{eq:chi2}
\end{equation}
For each FE-DVR grid except for the boundaries, the orthonormal functions $\chi _{i,\alpha}$ are obtained by normalizing the local basis function $\{f_{i,\alpha}\}$ as 
\begin{equation}
\chi _{i,\alpha}(q) =\frac{f_{i,\alpha}(q)}{\sqrt{w^{i}_{\alpha}}}.
\label{eq:chi3}
\end{equation}
The set of the functions in Eqs.~\eqref{eq:chi1}-\eqref{eq:chi3} satisfies the normalization condition in Eq.~\eqref{eq:norm_dvr}

\subsubsection{Implementation of the FE-DVR method}

By expanding the eigenfunction $\Psi (q)$ in terms of the FE-DVR functions $\chi _{i,\alpha}$, as shown in Eq.~\eqref{eq:psi}, the original equation can be mapped onto the eigenvalue problem. Since $\chi _{i,\alpha}$ is localized at each FE-DVR grid, the potential term $V(q)$ is diagonalized as 
\begin{equation}
\int^{R_R}_{R_L} \chi _{i,\alpha}(q) V(q) \chi _{j,\beta} (q) = \delta _{i,j}\delta _{\alpha,\beta}V(q^{i}_{\alpha}).
\end{equation} 

The matrix elements of the second derivative term for $i= 2, \dots, N-1$ are evaluated in terms of the FE-DVR basis as
\begin{align}
&\int^{R_R}_{R_L} \chi _{i,\alpha}(q) \left( - \frac{d^2}{dq^2} \right)
\chi _{j,\beta} (q) dq \nonumber \\
&=\mathcal{D}^i_{\alpha\beta}\delta _{i,j} 
+ \mathcal{L}^i_{\alpha,M}\delta _{i,j+1}\delta _{\beta,M}
+ \mathcal{U}^i_{M,\beta}\delta _{i,j-1}\delta _{\alpha,M} \nonumber \\
& + \delta _{i,1}\delta _{j,N}\delta _{\beta,M}\tilde{\mathcal{L}}_{\alpha}
+ \delta _{i,N}\delta _{j,1}\delta _{\alpha,M}\tilde{\mathcal{U}}_{\beta}, 
\end{align}
where the diagonal elements for $\alpha=\beta = M$ are given by 
\begin{align}
\mathcal{D}^i_{\alpha,\beta} = \sum^{M}_{\beta=1}
\bigg[ &
\frac{w^{i}_{\gamma}}{W^{i}_{M}}
f^{\prime}_{i,M}(q^{i}_{\gamma}) f^{\prime}_{i,M}(q^{i}_{\gamma}) \nonumber \\
&+ \frac{w^{i+1}_{\gamma}}{W^{i}_{M}}
f^{\prime}_{i+1,1}(q^{i+1}_{\gamma}) f^{\prime}_{i+1,1}(q^{i+1}_{\gamma})
\bigg],
\end{align}
otherwise 
\begin{equation}
\mathcal{D}^i_{\alpha,\beta} = \frac{w^{i}_{\beta}}{\sqrt{W^{i}_{\alpha}W^{i}_{\beta}}}
f^{\prime}_{i,\alpha}(q^{i}_{\alpha^{\prime}}) f^{\prime}_{i,\beta}(q^{i}_{\beta^{\prime}}).
\end{equation}
The other elements are 
\begin{align}
\mathcal{L}^{i\neq 1}_{\alpha,M} = \sum^{M}_{\gamma=1}
\frac{w^{i}_{\gamma}}{\sqrt{W^{i}_{\alpha}W^{i-1}_{M}}}
f^{\prime}_{i,\alpha}(q^{i}_{\gamma})f^{\prime}_{i,1}(q^{i}_{\beta}), 
\end{align}
\begin{align}
\mathcal{U}^{i\neq N}_{M,\beta} =
\sum^M_{\gamma=1}\frac{w^{i+1}_{\gamma}}{\sqrt{W^{i}_{M}W^{i+1}_{\beta}}}
f^{\prime}_{i+1,1}(q^{i+1}_{\gamma})f^{\prime}_{i+1,\beta}(q^{i+1}_{\gamma}).
\end{align}
and $\mathcal{L}^{i= 1}_{\alpha, M} = 0$ and $\mathcal{U}^{i= N}_{M, \beta} = 0$. The elements $\tilde{L}_{\alpha}$ and $\tilde{U}_{\beta}$ are obtained with the phase factor in Eq.~\eqref{eq:phase_factor} as
\begin{gather}
\tilde{\mathcal{L}}_{\alpha} = \mathcal{S}\sum^M_{\gamma = 1} \frac{w^1_{\gamma}}{\sqrt{W^1_{\alpha}W^N_M}}
f^{\prime}_{1,\alpha}(q^{1}_{\gamma})f^{\prime}_{1,1}(q^{1}_{\gamma}), \\
\tilde{\mathcal{U}}_{\beta} = \mathcal{S}^{\ast}
\sum^M_{\gamma = 1} \frac{w^1_{\gamma}}{\sqrt{W^N_MW^1_{\beta}}}
f^{\prime}_{1,1}(q^{1}_{\gamma})f^{\prime}_{1,\beta}(q^{1}_{\gamma}).
\end{gather}
We here set $W^{i}_{\alpha}= w^N_M+w^1_1$ for $(i,\alpha) = (N,M)$ and $(1,1)$ and $W^i_{\alpha}=w^{i}_{M}+w^{i+1}_1$ for $i\neq 1, N$ and $\alpha =M$, otherwise $W^{i}_{\alpha}=w^{i}_{\alpha}$.
The first derivatives of the local basis functions $f_{i,\alpha}(q)$ at quadrature points, 
$f^{\prime}_{i,\alpha}(q^{i}_{\beta}) \equiv
df_{i,\alpha}(q)/dq |_{q=q^{i}_{\beta}}$, are given by 
\begin{align}
f^{\prime}_{i,\alpha}(q^{i}_{\beta}) 
=
\left\{
\begin{array}{ll}
\displaystyle{
\frac{1}{q^{i}_{\alpha}-q^{i}_{\beta}}
\prod _{\gamma\neq\alpha,\beta} \frac{q^{i}_{\beta}-q^{i}_{\gamma}}{q^{i}_{\alpha}-q^{i}_{\gamma}}},
& \mbox{for $\alpha \!\neq\! \beta$}; \\
\\
\displaystyle{
\frac{1}{2w^{i}_{\alpha}} (\delta _{\alpha,M} - \delta _{\alpha,1})}, 
& \mbox{for $\alpha \!=\! \beta$}.
\end{array}
\right.
\end{align}

In the FE-DVR method, non-zero Hamiltonian matrix elements are composed of only the gradient terms and the potential terms is fully diagonalized. The number of non-zero elements, $N_\mathrm{mat}$, are determined by the number of DVR grids in each interval, $M$;  $N_\mathrm{mat} \sim M^2$. This drastically reduces the number of non-zero elements in the standard DVR method~\cite{dvr1,dvr2,dvr3} with $N_\mathrm{mat} \sim (NM)^2$. Hence, the FE-DVR method is accompanied by the sparse matrix and convenient for the implementation of the Lanczos/Arnoldi algorithm and  massive parallel computation.~\cite{parallel} 




\end{document}